\documentclass[10pt,reqno,a4paper]{amsart}
\usepackage[british]{babel}
\usepackage{amssymb,amstext,amsthm,eucal,bbm,mathrsfs,amscd,bbold}
\usepackage[margin=2.5cm]{geometry}
\usepackage{stmaryrd}
\usepackage{array}
\usepackage[svgnames]{xcolor}
\usepackage[utf8]{inputenc}
\usepackage{enumitem}
\usepackage[small]{eulervm}
\usepackage{tgpagella}
\usepackage{tikz,pgfplots}
\usetikzlibrary{calc}
\usepackage[unicode]{hyperref}
\hypersetup{%
  pdftitle   = {Kinematical Lie algebras in 2+1 dimensions},
  pdfkeywords = {Lie algebra, deformations, Galilean, Carroll, Newton, Poincaré},
  pdfauthor  = {Tomasz Andrzejewski, José Figueroa-O'Farrill},
  pdfcreator = {\LaTeX\ with package \flqq hyperref\frqq},
  linkcolor=NavyBlue,
  citecolor=ForestGreen,
  urlcolor=OrangeRed,
  anchorcolor=OrangeRed,
  colorlinks=true
}
\theoremstyle{plain}

\theoremstyle{definition}

\renewcommand{\a}{\boldsymbol{a}}
\renewcommand{\b}{\boldsymbol{b}}
\renewcommand{\c}{\boldsymbol{c}}
\newcommand{\aba}{\bar{\a}}
\newcommand{\bba}{\bar{\b}}
\newcommand{\cba}{\bar{\c}}
\renewcommand{\d}{\partial}
\newcommand{\g}{\mathfrak{g}}

\newcommand{\p}{\mathfrak{p}}
\newcommand{\co}{\mathfrak{co}}
\newcommand{\e}{\mathfrak{e}}
\renewcommand{\k}{\mathfrak{k}}

\newcommand{\so}{\mathfrak{so}}
\newcommand{\h}{\mathfrak{h}}
\renewcommand{\r}{\mathfrak{r}}

\renewcommand{\t}{\boldsymbol{t}}

\newcommand{\B}{\boldsymbol{B}}

\newcommand{\bbeta}{\boldsymbol{\beta}}
\newcommand{\bpi}{\boldsymbol{\pi}}
\renewcommand{\P}{\boldsymbol{P}}

\renewcommand{\H}{\mathcal{H}}

\newcommand{\ad}{\operatorname{ad}}
\newcommand{\id}{\mathbb{1}}

\newcommand{\RR}{\mathbb{R}}

\newcommand{\CC}{\mathbb{C}}

\newcommand{\GL}{\operatorname{GL}}

\newcommand{\CO}{\operatorname{CO}}
\newcommand{\Tr}{\operatorname{Tr}}

\renewcommand{\Re}{\operatorname{Re}}
\renewcommand{\Im}{\operatorname{Im}}

\newcommand{\Bbar}{\bar{\B}}
\newcommand{\Pbar}{\bar{\P}}
\newcommand{\gammabar}{\bar{\gamma}}
\newcommand{\betabar}{\bar{\bbeta}}
\newcommand{\pibar}{\bar{\bpi}}
\newcommand{\abar}{\bar{a}}
\newcommand{\bbar}{\bar{b}}
\newcommand{\cbar}{\bar{c}}
\newcommand{\dbar}{\bar{d}}

\newcommand{\Ebar}{\bar{E}}
\newcommand{\Deltabar}{\bar{\Delta}}
\definecolor{gris}{rgb}{0.5,0.5,0.5}
\newcommand{\zero}{{\color{gris}0}}
\allowdisplaybreaks[0]
\begin{document}

\title{Kinematical Lie algebras in $2+1$ dimensions}
\author[Andrzejewski]{Tomasz Andrzejewski}
\author[Figueroa-O'Farrill]{José Miguel Figueroa-O'Farrill}
\address[TA]{School of Physics and Astronomy, The University
  of Edinburgh, James Clerk Maxwell Building, Peter Guthrie Tait Road,
  Edinburgh EH9 3FD, United Kingdom}
\address[JMF]{Maxwell Institute and School of Mathematics, The University
  of Edinburgh, James Clerk Maxwell Building, Peter Guthrie Tait Road,
  Edinburgh EH9 3FD, United Kingdom}
\begin{abstract}
  We classify kinematical Lie algebras in dimension $2+1$.  This is
  approached via the classification of deformations of the static
  kinematical Lie algebra.  In addition, we determine which
  kinematical Lie algebras admit invariant symmetric inner products.
\end{abstract}
\thanks{EMPG-17-13}
\maketitle
\tableofcontents

\section{Introduction}
\label{sec:introduction}

One consequence of the principle of relativity, which from a purely
mathematical standpoint can be considered an instance of Klein's
Erlanger Programme, is that the geometry of the universe is dictated
by its Lie group of automorphisms.  As in Klein's programme, by
geometry one does not necessarily mean a metric geometry, but any sort
of geometrical datum which the automorphisms leave invariant.  In the
context of relativity, for example, the Newtonian model of the
universe, as an affine bundle (with three-dimensional fibres) over an
affine line, has the galilean group as automorphisms and the invariant
notions are time intervals between events and the euclidean distance
between simultaneous events.  By contrast, Minkowski spacetime has the
Poincaré group as the group of automorphisms and the invariant notion
is the proper distance (or, equivalently, the proper time).  Both the
galilean and Poincaré groups are examples of kinematical Lie groups,
whose Lie algebras (in dimension $2+1$) are the subject of this paper.

By a \textbf{kinematical Lie algebra} in dimension $D$, we mean a real
$\tfrac12 (D+1)(D+2)$-dimensional Lie algebra with generators
$R_{ab} = - R_{ba}$, with $1\leq a,b \leq D$, spanning a Lie
subalgebra isomorphic to $\so(D)$:
\begin{equation}
  [R_{ab}, R_{cd}] = \delta_{bc} R_{ad} -  \delta_{ac} R_{bd} -
  \delta_{bd} R_{ac} +  \delta_{ad} R_{bc},
\end{equation}
and $B_a$, $P_a$ and $H$ which transform according to the vector,
vector and scalar representations of $\so(D)$, respectively -- namely,
\begin{equation}
  \begin{split}
    [R_{ab}, B_c] &= \delta_{bc} B_a - \delta_{ac} B_b\\
    [R_{ab}, P_c] &= \delta_{bc} P_a - \delta_{ac} P_b\\
    [R_{ab}, H] &= 0.
  \end{split}
\end{equation}
The rest of the brackets between $B_a$, $P_a$ and $H$ are only subject
to the Jacobi identity: in particular, they must be
$\so(D)$-equivariant.  The kinematical Lie algebra where those
additional Lie brackets vanish is called the \textbf{static}
kinematical Lie algebra, of which, by definition, every other
kinematical Lie algebra is a deformation.

Up to isomorphism, there is only one kinematical Lie algebra in $D=0$:
it is one-dimensional and hence abelian.  For $D=1$, there are no
rotations and hence any three-dimensional Lie algebra is kinematical.
The classification is therefore the same as the celebrated Bianchi
classification of three-dimensional real Lie algebras \cite{Bianchi}.
The classification for $D=3$ is due to Bacry and Nuyts \cite{MR857383}
who completed earlier work of Bacry and Lévy-Leblond \cite{MR0238545}.
A deformation theory approach to the classification is described in
\cite{JMFKinematical3D}, which completes earlier work
\cite{JMFGalilean} for the galilean and Bargmann algebras, and which
also contains the classification of deformations of the universal
central extension of the static kinematical Lie algebra.  This
approach is used in \cite{JMFKinematicalHD} to classify the
kinematical Lie algebras for $D>3$ with and without central extension.
The purpose of this paper is to solve the classification problem for
$D=2$.  This problem is technically more involved than the problem for
higher $D$ for the simple reason that the representation of $\so(D)$
on $\RR^D$ has a larger endomorphism ring for $D=2$ than it does for
any $D>2$.  Indeed, despite being a real irreducible representation,
its endomorphism ring is the complex numbers.  This means that it is
often convenient to work not with real Lie algebras as for $D\geq 3$,
but with \emph{complexifications} of real Lie algebras; that is,
complex Lie algebras with real structures.  In order to trace a path
of least effort, we will freely move from one description to another
in this paper.  Sufficient information is given to allow the reader to
translate to their favourite formalism.

Let us remark in passing that the universal central extension of the
static kinematical Lie algebra in $D=2$ is also larger than in $D\geq
3$.  Whereas in $D\geq 3$ there is a one-dimensional central subspace,
in $D=2$ there is a five-dimensional central subspace spanned by
$Z_1,\dots,Z_5$ and brackets:
\begin{equation}
  [B_a, B_b] = \epsilon_{ab} Z_1 \qquad [B_a, P_b] = \delta_{ab} Z_2 +
  \epsilon_{ab} Z_3 \qquad [P_a, P_b] = \epsilon_{ab} Z_4
  \qquad\text{and}\qquad [R_{ab},H] = \epsilon_{ab} Z_5.
\end{equation}
The deformation problem of this centrally extended Lie algebra, while
potentially interesting, is beyond the scope of this paper.

We refer to \cite{JMFKinematical3D} for details on the methodology and
for a brief review of the basic notions of deformation theory and Lie
algebra cohomology, following \cite{MR0214636},
\cite{ChevalleyEilenberg} and \cite{MR0054581}.  In this approach, we
describe a Lie algebra structure on a vector space $V$ as an element
$\mu_0 \in \Lambda^2V^*\otimes V$ which has vanishing
Nijenhuis--Richardson bracket with itself $[\![\mu_0,\mu_0]\!]$.  This
bracket gives $L^\bullet := \Lambda^{\bullet + 1} V^*\otimes V$ the
structure of a graded Lie superalgebra.  In particular, the component
$[\![-,-]\!]: L^1 \times L^1 \to L^2$ of the bracket is symmetric.
Any other Lie algebra structure $\g = (V,\mu)$ defines $\varphi = \mu
- \mu_0 \in L^1$ which satisfies the Maurer--Cartan equation:
\begin{equation}
  \label{eq:maurer-cartan}
  \d \varphi  = \tfrac12 [\![\varphi,\varphi]\!],
\end{equation}
where $\d\varphi := -[\![\mu_0,\varphi]\!]$ is one component of the
Chevalley--Eilenberg differential of the Lie algebra
$\g_0 = (V,\mu_0)$ with values in the adjoint representation.  The
deformation theory approach is to solve the Maurer--Cartan equation
perturbatively, by writing $\varphi$ as a formal power series
$\varphi = \sum_{n=1}^\infty t^n \varphi_n$ and solving
equation~\eqref{eq:maurer-cartan} order by order in $t$.  The first
order equation says that $\d\varphi_1=0$.  We call such cocycles
\textbf{infinitesimal deformations} and each such $\varphi_1$ defines
a class in $H^2(\g_0;\g_0)$.  If this class is zero, then $\varphi_1$
is tangent to the $\GL(V)$ orbit of $\mu_0$ and we say that the
infinitesimal deformation is ineffective.  Therefore the interesting
infinitesimal deformations are those which are not cohomologically
trivial.  In practice we parametrise the space of infinitesimal
deformations by splitting the cohomology sequence
$B^2 \to Z^2 \to H^2(\g_0;\g_0)$ and choosing a convenient complement
$\H^2 \subset Z^2$ to the coboundaries $B^2$ in the space $Z^2$ of
cocycles.  The higher order terms in the Maurer--Cartan equation
\eqref{eq:maurer-cartan} can be understood as a sequence of
obstructions to integrating the infinitesimal deformation $\varphi_1$.
At every order in the perturbation expansion of the Maurer--Cartan
equation we find a cohomology class in $H^3$, whose vanishing is a
condition \emph{sine qua non} to be able to continue integrating the
deformation.  Although this process could in principle continue
indefinitely, it seldom does and indeed the deformations in this paper
are either obstructed or integrable at second order in the
perturbative expansion.

In this paper we are interested only in deformations of the static
kinematical Lie algebra which are themselves kinematical: i.e., such
that the Lie brackets $[R,-]$ involving the rotational generator
$R := -\frac12 \epsilon_{ab} R_{ab}$ are not modified or,
equivalently, that the deformation $\varphi$ obeys $\varphi(R,-) = 0$.
In other words, if we let $\r$ denote the Lie subalgebra of the static
kinematical Lie algebra spanned by $R$ and $\h$ the complementary
ideal spanned by $B_a$, $P_a$ and $H$, then\footnote{We will denote
  the static Lie algebra by $\g$, rather
  than $\g_0$, in an effort not to overburden ourselves
  notationally.} $\varphi \in \Lambda^2\h^* \otimes \g$. In other
words, the relevant deformation complex is the relative subcomplex
$C^\bullet(\g,\r;\g)$ which consists of the $\r$-invariant cochains in
$\Lambda^\bullet \h^*\otimes \g$. For $D\geq 3$, the relative
subcomplex is quasi-isomorphic to the full deformation complex, as a
consequence of the Hochschild--Serre decomposition theorem
\cite{MR0054581}, and therefore all deformations of the static
kinematical Lie algebra are automatically kinematical themselves. This
theorem is not applicable for $D=2$ because $\r$ is not semisimple
here. As a result there are in principle deformations which are not
kinematical. (In fact, the space of all infinitesimal deformations is
$19$-dimensional, whereas as we will see the space corresponding to
kinematical deformations is ``only'' $11$-dimensional.)

An important characteristic of a Lie algebra, particularly for
applications in field theory, is whether or not the Lie algebra admits
a symmetric inner product which is invariant under the adjoint
action of the Lie algebra on itself.  Such Lie algebras are said to be
\emph{metric}.  In this paper we also determine which kinematical
Lie algebras are metric.  We will see that similar to what happens in
$D=3$ and contrary to what happens in dimension $D>3$, there are
non-semisimple metric kinematical Lie algebras.

The plan of this paper is the following.  In
Section~\ref{sec:def-comp} we describe the deformation complex, but we
relegate to Appendix~\ref{app:enum-complex} the precise enumeration of
cochains that we will use in our calculations, as well as the relevant
component of the Nijenhuis--Richardson bracket.  There are two
complementary descriptions of kinematical Lie algebras in this
dimension: one is as real Lie algebras and the other as complex Lie
algebras with a real structure.  This second description simplifies
the discussion of automorphisms, which will play a crucial rôle in
this approach.  In Section~\ref{sec:automorphisms} we describe the
group $G$ of automorphisms of $\g$ which preserve the deformation
complex.  This will play an important rôle when we split the cohomology
sequence to parametrise the space of infinitesimal deformations, when
we solve the obstruction relations and also when we classify the
different integrable deformations up to isomorphism. In
Section~\ref{sec:infin-defs} we calculate the second cohomology of the
deformation complex and choose a convenient $G$-stable parametrisation
of the infinitesimal deformations, whose obstructions are analysed in
Section~\ref{sec:obstructions}.  We find that integrable deformations
are of at most second order and they fall into one of four branches
labelling the $G$-orbits in a four-dimensional subspace of the
cohomology.  In Section~\ref{sec:deformations} we study the
isomorphism classes of integrable deformations for each of those
branches.  The main technique is to exploit the stabiliser of the
typical point in each orbit to bring the remaining free parameters to
a canonical form.  Doing so for each orbit we arrive at the
classification which is summarised in Table~\ref{tab:summary} in
Section~\ref{sec:summary}, which also contains the information of
which deformations are metric, as determined in
Section~\ref{sec:invar-inner-prod}.

\section{The deformation complex}
\label{sec:def-comp}

Let $\g$ be the static kinematical Lie algebra for $D=2$.  It is
spanned by $R,B_a, P_a, H$ subject to the following nonzero Lie
brackets:
\begin{equation}
  \label{eq:static-R}
  [R,B_a] = \epsilon_{ab} B_b \qquad\text{and}\qquad
  [R,P_a] = \epsilon_{ab} P_b.
\end{equation}
Let $\r \cong \so(2)$ denote the abelian Lie subalgebra spanned by $R$
and let $\h$ denote the abelian ideal spanned by $B_a,P_a,H$.  Let
$\beta_a, \pi_a, \eta$ denote the canonical dual basis for $\h^*$.

We may diagonalise the action of $R$ by complexifying $\g$.  This will
turn out to simplify the action of automorphisms on the deformation
complex, so we will also describe this approach.  To this end we
introduce $\B = B_1 + i B_2$ and $\P = P_1 + i P_2$ and extend the Lie
brackets complex-linearly, so that now
\begin{equation}
  \label{eq:static-C}
  [R,\B] = - i \B \qquad\text{and}\qquad [R, \P] = -i \P.
\end{equation}
We also have $\Bbar = B_1 - i B_2$ and $\Pbar = P_1 - i P_2$, which
satisfy
\begin{equation}
  [R,\Bbar] = i \Bbar \qquad\text{and}\qquad [R,\Pbar] = i \Pbar.
\end{equation}
The \emph{complex} span of $R,H,\B,\P,\Bbar,\Pbar$, which we denote by
$\CC\left<R,H,\B,\P,\Bbar,\Pbar\right>$, defines a complex Lie algebra
$\g_\CC$.  This complex Lie algebra has a conjugation (that is, a
complex-antilinear involutive automorphism) denoted by $\star$ and
defined by $H^\star = H$, $R^\star = R$, $\B^\star = \Bbar$ and
$\P^\star = \Pbar$.  We see that the real Lie subalgebra of $\g_\CC$
consisting of real elements (i.e., those $X \in \g_\CC$ such that
$X^\star = X$) is the static kinematical Lie algebra $\g$.

Let $\h_\CC$ denote the ideal of $\g_\CC$ spanned by
$H,\B,\P,\Bbar,\Pbar$.  We will let $\eta, \bbeta, \bpi, \betabar, \pibar$
denote the canonical dual basis for $\h^*_\CC$.  These are related
to $\beta_a$ and $\pi_a$ by the following relations:
\begin{equation}
  \bbeta = \tfrac12 (\beta_1 - i \beta_2) \qquad\text{and}\qquad \bpi =
  \tfrac12 (\pi_1 - i \pi_2),
\end{equation}
with $\betabar$ and $\pibar$ being their naive complex conjugates.
We extend the action of the conjugation $\star$ to $\h_\CC^*$ by
$\eta^\star = \eta$, $\bbeta^\star = \betabar$ and $\bpi^\star =
\pibar$.

We are interested in kinematical Lie algebras, so we are not deforming
the Lie brackets involving the rotation generator; that is, $B_a$ and
$P_a$ still transform as vectors and $H$ still transforms as a
scalar.  The Jacobi identity then says that the Lie brackets must be
$\r$-equivariant.  This implies that the deformation complex
is thus $C^\bullet := C^\bullet(\g,\r;\g) \cong \left(\Lambda^\bullet \h^*
\otimes \g\right)^\r$, which can also be identified with the
$\r$-invariant subcomplex of the Chevalley--Eilenberg complex of the
abelian Lie algebra $\h$ with values in the representation $\g$.

The deformation complex $C^\bullet$ can also be identified with the
real subcomplex of
$C^\bullet_\CC := \left(\Lambda^\bullet \h^*_\CC \otimes
  \g_\CC\right)^\r$.  This real subcomplex consists of those cochains
which are fixed by the conjugation $\star$.  At a practical level, one
can work with $C^\bullet$ by working with $C^\bullet_\CC$ and making
sure that one considers only real elements.  This turns out to be very
convenient when discussing automorphisms, since these act more simply
and more naturally on $C^\bullet_\CC$.

The real dimension of $C^\bullet$ is the complex dimension of
$C^\bullet_\CC$, which can be calculated using character theory as
follows.  The character $\chi_{\g_\CC}(q)$ of $\g_\CC$ as a
representation of $\r$ is given by
\begin{equation}
  \chi_{\g_\CC}(q) = 2 + 2 (q + q^{-1}),
\end{equation}
whereas that of $\h_\CC$ is given by
\begin{equation}
  \chi_{\h_\CC}(q) = 1 + 2 (q + q^{-1}).
\end{equation}
Since this is invariant under $q \mapsto q^{-1}$, this is also the
character $\chi_{\h_\CC^*}$ of $\h_\CC^*$.  The character of
$\Lambda^p \h_\CC^* \otimes \g_\CC$ can be calculated as follows.
First of all, since characters are multiplicative over the tensor
product,
\begin{equation}
 \chi_{\Lambda^p\h_\CC^* \otimes \g_\CC}(q) =   \chi_{\Lambda^p\h_\CC^*}(q)
  \chi_{\g_\CC}(q) =  2(1 + q + q^{-1}) \chi_{\Lambda^p\h_\CC}(q).
\end{equation}
The character for the $\Lambda^p\h_\CC$ can be read off from their
generating function:
\begin{equation}
  \sum_{n=0}^\infty t^n \chi_{\Lambda^n\h_\CC}(q) = \exp \left(-
    \sum_{\ell=1}^\infty\frac{(-t)^\ell}{\ell} \chi_{\h_\CC}(q^\ell) \right).
\end{equation}
Expanding this to second order we find that $\chi_{\Lambda^0\h_\CC}(q) =
1$, $\chi_{\Lambda^1\h_\CC}(q) = \chi_{\h_\CC}(q)$ and
\begin{equation}
  \chi_{\Lambda^2\h_\CC}(q) = \tfrac12 \left(\chi_{\h_\CC}(q)^2 - \chi_{\h_\CC}(q^2)\right) = q^{-2} + 2 q^{-1} + 4 + 2 q + q^2,
\end{equation}
  and by Poincaré duality $\chi_{\Lambda^3\h_\CC}(q) = \chi_{\Lambda^2\h_\CC}(q)$.  Therefore,
\begin{equation}
  \begin{split}
    \chi_{\Lambda^0\h_\CC^* \otimes \g_\CC}(q) &= 2 q^{-1} + 2 + 2 q \implies \dim C^0 = 2\\
    \chi_{\Lambda^1\h_\CC^* \otimes \g_\CC}(q) &=4 q^{-2} + 6 q^{-1} + 10 + 6 q + 4 q^2 \implies \dim C^1 = 10\\
    \chi_{\Lambda^2\h_\CC^* \otimes \g_\CC}(q) &=2 q^{-3} + 6 q^{-2} + 14 q^{-1} + 16 + 14 q + 6 q^2 + 2 q^3 \implies  \dim C^2 = 16,
  \end{split}
\end{equation}
and again $\dim C^3 = 16$ by duality.

In Appendix~\ref{app:enum-complex} we define bases for the
$C^0,\dots,C^3$ and $C^0_\CC,\dots,C^3_\CC$, as well as a dictionary
between the two bases.  We also tabulate the Nijenhuis--Richardson
product on the space of 2-cochains, which will be useful when
computing the obstructions to infinitesimal deformations.

The Chevalley--Eilenberg differential on $C^\bullet$ is defined on
generators by
\begin{equation}
  \label{eq:CE}
  \d R = -\epsilon_{ab} (\beta_a B_b + \pi_a P_b)
  \qquad \d \beta_a = \d \pi_a = \d \eta = \d B_a = \d P_a = \d H = 0,
\end{equation}
and the one on $C^\bullet_\CC$ is given by
\begin{equation}
  \label{eq:CE-C}
  \d R = i\bbeta\B - i\betabar\Bbar + i \bpi\P - i \pibar\Pbar
 \qquad\text{and}\qquad \d \bbeta = \d \bpi = \d \eta = \d\B = \d\P = \d H = 0.
\end{equation}
The differential is real, so that $\d \Bbar = \overline{\d \B} =0$, et
cetera, so the real elements of $C^\bullet_\CC$ do indeed form a
subcomplex.  From the above formulae it is easy to calculate the
differential on the bases given in Appendix~\ref{app:enum-complex}.

\section{Automorphisms}
\label{sec:automorphisms}

For the static kinematical Lie algebra in dimension $D\geq 3$, the
subgroup of automorphisms of $\g$ which preserves the deformation
complex is $\GL(\RR^2) \times \RR^\times$ (see, e.g.,
\cite{JMFKinematical3D,JMFKinematicalHD}).  For $D=2$ this is enhanced
to $\GL(\CC^2) \times \RR^\times$.  This is transparent in the complex
version of the Lie algebra, where the action of $G$ on the generators
of $\g_\CC$ is given by declaring $R$ to be invariant and by
\begin{equation}
  (\B, \P, H) \mapsto (\B, \P, H)
  \begin{pmatrix}
    a & b & \zero \\ c & d & \zero \\ \zero & \zero & \lambda
  \end{pmatrix}
  \qquad\text{where}\quad
  \begin{pmatrix}
    a & b \\ c & d
  \end{pmatrix} \in \GL(\CC^2) \quad\text{and}\quad \lambda \in
  \RR^\times,
\end{equation}
with the induced action on the generators of $\h_\CC^*$ is given by the
transpose inverse:
\begin{equation}
  (\bbeta, \bpi, \eta) \mapsto (\bbeta, \bpi, \eta)
  \begin{pmatrix}
    d/\Delta & -c/\Delta & \zero \\
    -b/\Delta & a/\Delta & \zero \\
    \zero & \zero & \lambda^{-1}
  \end{pmatrix}
  \qquad\text{where}\quad \Delta = ad - bc.
\end{equation}
In order to ensure that the automorphisms preserve the real
deformation complex, we must define the action of $G$ on
$\Bbar,\Pbar,\betabar,\pibar$ simply by complex conjugating the above
formulae.

In summary, and being more explicit,
\begin{equation}
  \label{eq:autos}
  \begin{aligned}[m]
    \B &\mapsto a \B + c \P \\
    \P &\mapsto b \B + d \P \\
    H & \mapsto \lambda H
  \end{aligned}
  \qquad\qquad
  \begin{aligned}[m]
    \bbeta & \mapsto \Delta^{-1} ( d \bbeta - b \bpi)\\
    \bpi & \mapsto \Delta^{-1} (-c \bbeta + a \bpi)\\
    \eta & \mapsto \lambda^{-1} \eta
  \end{aligned},
\end{equation}
with $R$ invariant and $\Bbar \mapsto \abar \Bbar + \cbar \Pbar$, et
cetera.

From this we can work out the action of $G$ on the bases given in
Appendix~\ref{sec:enum-compl-deform}.  For $C^1_\CC$ we find
\begin{equation}
  \label{eq:autos-C1}
  \begin{aligned}[m]
    \a_1 &\mapsto \lambda^{-1} \a_1\\
    \a_2 &\mapsto \a_2\\
    \a_3 + \a_6 &\mapsto \a_3 + \a_6\\
  \end{aligned}
  \qquad\qquad
  \begin{aligned}[m]
    \a_3 - \a_6 &\mapsto \Delta^{-1} ((ad+bc) (\a_3 - \a_6) + 2 cd \a_4 - 2ab \a_5)\\
    \a_4 &\mapsto \Delta^{-1} (bd (\a_3 - \a_6) + d^2 \a_4 - b^2 \a_5)\\
    \a_5 &\mapsto \Delta^{-1} (-ac (\a_3 - \a_6) - c^2 \a_4 + a^2 \a_5)\\
  \end{aligned}
\end{equation}
and for $C^2_\CC$ we find
\begin{equation}
  \label{eq:autos-C2}
  \begin{aligned}[m]
    \c_1 + \c_4 &\mapsto \lambda^{-1} (\c_1 + \c_4)\\
    \c_1 - \c_4 &\mapsto \lambda^{-1} \Delta^{-1} ( (ad + bc) (\c_1 -\c_4) + 2 cd \c_2 - 2 ab \c_3 )\\
    \c_2 &\mapsto \lambda^{-1}\Delta^{-1}( bd (\c_1 - \c_4) + d^2 \c_2 - b^2 \c_3 )\\
    \c_3 &\mapsto \lambda^{-1}\Delta^{-1}(-ac (\c_1 - \c_4) - c^2 \c_2 + a^2 \c_3 )\\
    \c_5 &\mapsto |\Delta|^{-2} (|d|^2 \c_5 - i\bbar d \c_7 + i  b\dbar \cba_7  + |b|^2 \c_9 )\\
    \c_7 &\mapsto |\Delta|^{-2} (i \cbar d \c_5 +\abar d \c_7 - b\cbar \cba_7 + i\abar b \c_9)\\
    \c_9 &\mapsto |\Delta|^{-2}(|c|^2 \c_5 -i \abar c \c_7 + i a \cbar \cba_7 + |a|^2 \c_9)\\
    \c_6 &\mapsto \lambda|\Delta|^{-2} (|d|^2 \c_6 - i\bbar d \c_8 + i b\dbar \cba_8  + |b|^2 \c_{10} )\\
    \c_8 &\mapsto \lambda|\Delta|^{-2} (i \cbar d \c_6 +\abar d \c_8 - b\cbar \cba_8 + i \abar b  \c_{10})\\
    \c_{10} &\mapsto \lambda|\Delta|^{-2}(|c|^2 \c_6 -i \abar c \c_8 + i a \cbar \cba_8 + |a|^2 \c_{10}).
  \end{aligned}
\end{equation}

Let us point out that the representation of $G$ on the
four-dimensional complex vector space with ordered basis $(\c_{10}, i\cba_8,
-i\c_8, \c_6)$ is such that $(A,\lambda) \in G$ acts via the matrix
$\lambda M_A$, where
\begin{equation}
  \label{eq:4drep-G}
  M_A := \frac1{|\Delta|^2}
  \begin{pmatrix}
    |a|^2 & a \bbar & b\abar & |b|^2\\
    a\cbar & a\dbar & b\cbar & b\dbar\\
    c\abar & c \bbar &d \abar & d \bbar\\
    |c|^2 & c \dbar & d \cbar & |d|^2
  \end{pmatrix} =
  \frac{1}{\Delta}
  \begin{pmatrix}
    a & b \\ c & d 
  \end{pmatrix} \otimes
  \frac{1}{\Deltabar}
  \begin{pmatrix}
    \abar & \bbar \\ \cbar & \dbar
  \end{pmatrix},
\end{equation}
which shows that the representation of $\GL(\CC^2)$ which sends
$A \mapsto M_A$ is isomorphic to
$(\Lambda^2E^* \otimes E) \otimes (\Lambda^2\Ebar^* \otimes
  \Ebar)$, where $E= \CC^2$ is the identity representation of
$\GL(\CC^2)$ and $\Ebar$ the conjugate representation.  In the
symmetric square of this representation there is a submodule
isomorphic to
$(\Lambda^2 E^*)^2\otimes \Lambda^2E \otimes (\Lambda^2 \Ebar^*)^2
\otimes \Lambda^2\Ebar \cong \Lambda^2 E^* \otimes \Lambda^2 \Ebar^*$
and this means that there is a symmetric bilinear form $K$ which obeys
\begin{equation}
  M_A^T K M_A =|\Delta|^{-2} K.
\end{equation}
Relative to the basis $(\c_{10}, i\cba_8, -i\c_8, \c_6)$, the matrix
$K$ is given (up to a scale) by
\begin{equation}
  \label{eq:lorentzian-ip}
  K =
  \begin{pmatrix}
    \zero & \zero & \zero & -1\\
    \zero & \zero & 1 & \zero \\
    \zero & 1 & \zero & \zero \\
    -1 & \zero & \zero & \zero \\
  \end{pmatrix}.
\end{equation}

The action on the real basis for $C^1$ and $C^2$ is more cumbersome
and we will not write it down.  Our strategy shall be that we will
calculate infinitesimal deformations and obstructions using the real
complex $C^\bullet$ and the real basis, but shall complexify to
$C^\bullet_\CC$ and use the complex basis when discussing the action
of automorphisms.

From \eqref{eq:autos-C2} it follows that $C^2_\CC$ decomposes into the
following complex $G$-submodules:
\begin{equation}
  \label{eq:C2-C-Gmod}
  C_\CC^2 =\CC\left<\c_1 + \c_4\right> \oplus \CC\left<\c_1-\c_4, \c_2,
    \c_3\right> \oplus \CC\left<\cba_1-\cba_4, \cba_2, \cba_3\right> \oplus
  \CC\left<\c_5,\c_7,\cba_7, \c_9\right> \oplus \CC \left<\c_6,\c_8, \cba_8,
    \c_{10}\right>,
\end{equation}
and, in turn, this decomposes the real subspace $C^2$ into the
following real $G$-submodules:
\begin{multline}
  \label{eq:C2-Gmod-C}
  C^2 = \RR\left<\Re(\c_1 + \c_4)\right> \oplus \RR\left<\Im(\c_1 +
    \c_4)\right> \oplus \RR\left<\Re(\c_1-\c_4), \Im(\c_1-\c_4),
    \Re\c_2,\Im\c_2, \Re\c_3,\Im\c_3\right>\\
  \oplus \RR\left<\c_5,\Re\c_7,\Im\c_7, \c_9\right> \oplus \RR
  \left<\c_6,\Re\c_8, \Im\c_8, \c_{10}\right>,
\end{multline}
or in terms of the real basis,
\begin{multline}
    \label{eq:C2-Gmod-R}
  C^2 = \RR\left<c_1 + c_7\right> \oplus \RR \left<c_2 + c_8\right>
  \oplus \RR \left<c_1-c_7, c_2-c_8, c_3, c_4, c_5, c_6\right>\\ \oplus
  \RR\left<c_9, c_{11}, c_{12}, c_{15}\right> \oplus \RR \left<c_{10},
  c_{13}, c_{14}, c_{16}\right>.
\end{multline}

\section{Infinitesimal deformations}
\label{sec:infin-defs}

Infinitesimal (kinematical) deformations of the static Lie algebra
$\g$ are classified by $H^2(\g,\r;\g)$, which as explained above is
isomorphic to $H^2(\h;\g)^\r$.  From the expression of the
Chevalley--Eilenberg differential on generators given in
equation~\eqref{eq:CE}, we can compute the spaces of cocycles and
coboundaries in low degree.  Recall that $C^0$ is spanned by $R$ and
$H$.  Clearly $H$ is a cocycle, so $B^1$ is spanned by
$\d R = -a_4 - a_{10} = - 2\Im(\a_3+\a_6)$.  We see from
equation~\eqref{eq:autos-C1} that $\d R$ is indeed invariant under
$G$.  The differential $\d: C^1 \to C^2$ is given by
$\d a_1 = c_{2} + c_{8} = 2\Im(\c_1 + \c_4)$ and zero on the other
basis cochains.  Therefore $B^2$ is spanned by $\Im(\c_1 + \c_4)$,
which from \eqref{eq:autos-C2} we see that it is a $G$-submodule, as
expected.  The differential $\d: C^2 \to C^3$ is given by
\begin{equation}
  \d c_{9} = b_{14}, \quad
  \d c_{11} = \tfrac12 (b_{12} - b_9), \quad
  \d c_{12} = b_{16} - b_{13} \quad\text{and}\quad
  \d c_{15} = - b_{15},
\end{equation}
and zero on the other basis cochains.  Therefore,
\begin{equation}
  Z^2 = \RR \left<c_1,\dots,c_{8}, c_{10}, c_{13}, c_{14}, c_{16} \right>.
\end{equation}

We wish to split the sequence
\begin{equation}
  \begin{CD}
    0 @>>> B^2 @>>> Z^2 @>>> H^2 @>>> 0
  \end{CD}
\end{equation}
by choosing a subspace $\H^2 \subset Z^2$ which is stable under the
action of the group $G$ of automorphisms.  From the explicit
decomposition of $C^2$ as $G$-submodules in \eqref{eq:C2-Gmod-C}, we
find that the subspace $\H^2$ can be chosen to be the following direct
sum of $G$-submodules of $Z^2$:
\begin{multline}
  \H^2 = \RR\left<\Re(\c_1 + \c_4)\right> \oplus \RR\left<\Re(\c_1-\c_4), \Im(\c_1-\c_4),
    \Re\c_2,\Im\c_2, \Re\c_3,\Im\c_3\right>\\ \oplus \RR
  \left<\c_6,\Re\c_8, \Im\c_8, \c_{10}\right>,
\end{multline}
or in terms of the real basis
\begin{equation}
  \label{eq:H2-Gmod}
  \H^2 = \RR \left<c_1 + c_7\right> \oplus \RR\left<c_1 - c_7, c_2 -
    c_8, c_3, c_4, c_5, c_6\right> \oplus
  \RR\left<c_{10},c_{13},c_{14},c_{16}\right>.
\end{equation}

We therefore have an 11-dimensional space of infinitesimal deformations, parametrised as:
\begin{equation}
  \label{eq:infi-def}
  \varphi_1 = t_1 (c_{1}+c_{7}) + t_2 (c_{1}-c_{7}) + t_3 c_{3} + t_4
  c_{5} + t_5 (c_{2}-c_{8}) + t_6 c_{4} + t_7 c_{6} +t_8 c_{14} +
  t_9 c_{10} + t_{10} c_{16} + t_{11} c_{13},
\end{equation}
where the order has been chosen for later computational convenience.

We claim that the action of $G$ on the four-dimensional space of
infinitesimal deformations parametrised by $t_8,t_9,t_{10},t_{11}$ is
essentially a four-dimensional Lorentz transformation and a dilation.
To see this, notice that this component of the deformation is given by
\begin{equation}
  t_8 c_{14} + t_9 c_{10} + t_{10} c_{16} + t_{11} c_{13} = 2
  \left(- t_{10} \c_{10} + (t_8 - i t_{11}) (-i\cba_8) + (t_8 + i
    t_{11}) (-i \c_8) - t_9 \c_6  \right),
\end{equation}
using the dictionary in equation~\eqref{eq:dictionary-C2}.  As shown
in Section~\ref{sec:automorphisms}, the action of $G$ preserves the
conformal class of the inner product defined by $K$ in equation
\eqref{eq:lorentzian-ip}.  The norm of
$t_8 c_{14} + t_9 c_{10} + t_{10} c_{16} + t_{11} c_{13}$ relative to
that inner product is (up to an inconsequential factor):
\begin{equation}
  (-t_{10}, t_8 - i t_{11}, t_8 + i t_{11}, -t_9) K   (-t_{10}, t_8 - i
  t_{11}, t_8 + i t_{11}, -t_9)^T = t_8^2 - t_9 t_{10} + t_{11}^2,
\end{equation}
which has lorentzian signature.  There are four $G$-orbits in that
four-dimensional vector space, labelled by the following choices for
the vector  $\t = (t_8,t_9,t_{10},t_{11})$:
\begin{enumerate}
\item the \textbf{zero orbit} of the vector $\t=(0,0,0,0)$;
\item the \textbf{lightlike orbit} of the vector $\t=(0,1,0,0)$;
\item the \textbf{timelike orbit} of the vector $\t=(0,1,1,0)$; and
\item the \textbf{spacelike orbit} of the vector $\t=(1,0,0,0)$.
\end{enumerate}

Let us now consider the obstructions to integrating the infinitesimal
deformations found above.

\section{Obstructions}
\label{sec:obstructions}

The first obstruction is the class of
$\tfrac12 [\![\varphi_1,\varphi_1]\!]$ in $H^3$, which can be
calculated from the explicit expression~\eqref{eq:NR-bracket-C2} for
the Nijenhuis--Richardson bracket.  Its vanishing in cohomology is
equivalent to the following system of quadrics (after some
simplification):
\begin{equation}
  \label{eq:quadrics}
  \begin{aligned}[m]
    0&= 2 t_5 t_8 - t_7 t_9 + t_6 t_{10}\\
    0&= 2 t_5 t_{11} + t_4 t_9 + t_3 t_{10}\\
    0&= t_3 t_8 + t_2 t_9 - t_6 t_{11}\\
    0&= t_4 t_8 - t_2 t_{10} + t_7 t_{11}\\
  \end{aligned}
  \qquad\qquad
  \begin{aligned}[m]
    0&= t_1 t_8\\
    0&= t_1 t_9\\
    0&= t_1 t_{10}\\
    0&= t_1 t_{11}\\
  \end{aligned}
\end{equation}
Assuming these equations are satisfied, $\tfrac12
[\![\varphi_1,\varphi_1]\!] = \d \varphi_2$, where
\begin{equation}
\label{eq:def-2}
  \varphi_2 = (t_3 t_{11} + t_5 t_9 + t_6 t_8) c_{9} + (t_2 t_8 - t_4
  t_9 - t_5 t_{11}) c_{11} +  (-t_5 t_8 - t_2 t_{11} + t_7 t_9) c_{12}
  + ( t_7  t_8 - t_5 t_{10} - t_4 t_{11} ) c_{15}.
\end{equation}

The next obstruction is the class of $[\![\varphi_1,\varphi_2]\!]$ in
$H^3$, which again can be calculated from \eqref{eq:NR-bracket-C2}.
Demanding that this vanishes, we obtain a number of cubic equations,
which together with the quadrics leads to some simplification:
\begin{equation}
  \label{eq:cubics}
  \begin{split}
    0&= t_8 (2 t_2 t_5 + t_4 t_6 + t_3 t_7)\\
    0&= t_9 (2 t_2 t_5 + t_4 t_6 + t_3 t_7)\\
    0&= t_{10} (2 t_2 t_5 + t_4 t_6 + t_3 t_7)\\
    0&= t_{11} (2 t_2 t_5 + t_4 t_6 + t_3 t_7);
\end{split}
\end{equation}
although only three are independent once the quadrics are taken into
account.  If these cubic equations are satisfied, it is not just the
cohomology class of $[\![\varphi_1,\varphi_2]\!]$ which vanishes, but
the cocycle itself.  Therefore we can take $\varphi_3 = 0$.  Finally,
we see from \eqref{eq:NR-bracket-C2} that the cochains
$c_9,c_{11},c_{12},c_{15}$ appearing in $\varphi_2$ have vanishing
Nijenhuis--Richardson brackets among themselves, so that also
$[\![\varphi_2,\varphi_2]\!]=0$ and hence the deformation integrates
at second order.

In summary, we have the following deformation
\begin{multline}
  \label{eq:def}
  \varphi = t_1 (c_{1}+c_{7}) + t_2 (c_{1}-c_{7}) + t_3 c_{3} + t_4
  c_{5} + t_5 (c_{2}-c_{8}) + t_6 c_{4} + t_7 c_{6}\\ + t_8 c_{14} +
  t_9 c_{10} + t_{10} c_{16} + t_{11} c_{13} + (t_3 t_{11} + t_5 t_9 +
  t_6 t_8) c_{9} + (t_2 t_8 - t_4 t_9 - t_5 t_{11}) c_{11} \\ +  (t_7
  t_9 - t_2 t_{11} - t_5 t_8) c_{12} + ( t_7  t_8 - t_5 t_{10} - t_4 t_{11} ) c_{15}
\end{multline}
subject to the following integrability equations:
\begin{equation}
  \label{eq:relations}
  \begin{aligned}[m]
    0&= t_1 t_8\\
    0&= t_1 t_9\\
    0&= t_1 t_{10}\\
    0&= t_1 t_{11}\\
  \end{aligned}
  \qquad\qquad
  \begin{aligned}[m]
    0&= 2 t_5 t_8 - t_7 t_9 + t_6 t_{10}\\
    0&= 2 t_5 t_{11} + t_4 t_9 + t_3 t_{10}\\
    0&= t_3 t_8 + t_2 t_9 - t_6 t_{11}\\
    0&= t_4 t_8 - t_2 t_{10} + t_7 t_{11}\\
  \end{aligned}
  \qquad\qquad
  \begin{aligned}[m]
    0&= t_8 (2 t_2 t_5 + t_4 t_6 + t_3 t_7)\\
    0&= t_9 (2 t_2 t_5 + t_4 t_6 + t_3 t_7)\\
    0&= t_{10} (2 t_2 t_5 + t_4 t_6 + t_3 t_7)\\
    0&= t_{11} (2 t_2 t_5 + t_4 t_6 + t_3 t_7).
  \end{aligned}
\end{equation}

\section{Deformations}
\label{sec:deformations}

While it is possible to solve the obstruction relations \eqref{eq:relations}
using Gröbner methods, it is much more transparent to exploit the
automorphisms and in particular the orbit decomposition discussed at
the end of Section~\ref{sec:infin-defs}.  This leads us to consider
the four branches of solutions into which this section is divided.

\subsection{Zero orbit}
\label{sec:zero-orbit}

Here $t_8 = t_9 = t_{10} = t_{11} = 0$ and hence all obstruction
relations are satisfied.  Therefore we find that
for all $t_1,t_2,t_3,t_4,t_5,t_6,t_7$, the following
deformation is integrable:
\begin{equation}
  \begin{split}
    \varphi &= t_1 (c_{1}+c_{7}) + t_2 (c_{1}-c_{7}) + t_3 c_{3} + t_4
    c_{5} + t_5 (c_{2}-c_{8}) + t_6 c_{4} + t_7 c_{6}\\
    &= (t_1 + t_2 - it_5) \c_1 + (t_3 - it_6) \c_2 + (t_4 - i t_7)
    \c_3 + (t_1 - t_2 + it_5) \c_4 + \text{c.c.}
\end{split}
\end{equation}
The corresponding brackets are
\begin{equation}
  \begin{aligned}[m]
    [R,B_a] &= \epsilon_{ab} B_b\\
    [R,P_a] &= \epsilon_{ab} P_b\\
    \end{aligned}
    \qquad\qquad
    \begin{aligned}[m]
    [H,B_a] &= (t_1+t_2) B_a + t_5 \epsilon_{ab} B_b + t_3 P_a + t_6 \epsilon_{ab} P_b\\
    [H,P_a] &= t_4 B_a + t_7 \epsilon_{ab} B_b + (t_1-t_2) P_a - t_5 \epsilon_{ab} P_b,
  \end{aligned}
\end{equation}
or in complex form
\begin{equation}
  \begin{aligned}[m]
    [R,\B] &= -i \B\\
    [R,\P] &= -i \P\\
    \end{aligned}
    \qquad\qquad
    \begin{aligned}[m]
    [H,\B] &= (t_1 + t_2 - i t_5) \B + (t_3 - i t_6) \P\\
    [H,\P] &= (t_4 - i t_7) \B + (t_1 - t_2 + i t_5) \P.
  \end{aligned}
\end{equation}

We may now use $G$ (which we have not used yet, since the zero vector
has all of $G$ as stabiliser) to bring the bracket to a normal form.
Recall that $G$ acts as general linear transformations in $\B$ and
$\P$ and by rescaling $H$ by a nonzero real number.  The adjoint
action of $H$ on $\B$ and $\P$ is defined by the matrix
\begin{equation}
  M_H =
  \begin{pmatrix}
    t_1 + t_2 - i t_5 & t_4 - i t_7\\
    t_3 i t_6 & t_1 - t_2 + i t_5
  \end{pmatrix},
\end{equation}
so that under $G$
\begin{equation}
  M_H \mapsto \lambda A^{-1} M_H A \qquad\text{where}\qquad A =
  \begin{pmatrix}
    a & b \\ c & d
  \end{pmatrix} \in \GL(\CC^2)
\end{equation}
and $\lambda \in \RR^\times$.  In other words, we can conjugate $M_H$
and multiply it by a real scale.

Let us first focus on conjugation, which does not change the trace,
which we see from the explicit form of $M_H$ that it is real and equal
to $2 t_1$.  A complex $2\times 2$ matrix is either diagonalisable or
not.  If diagonalisable, it may be conjugated to a diagonal matrix,
which, if it has real trace, must take the form
\begin{equation}
  \label{eq:normal-form-diag}
  \begin{pmatrix}
    \mu_1 + i \theta & \zero \\ \zero & \mu_2 - i \theta
  \end{pmatrix}\qquad \text{for some $\mu_i,\theta \in \RR$.}
\end{equation}
Moreover, by relabelling $B$ and $P$, if necessary, we can assume that
$\mu_1 \geq \mu_2$.  If $M_H$ is not diagonalisable, then we can bring
it to a Jordan form,
\begin{equation}
  \label{eq:normal-form-jordan}
  \begin{pmatrix}
    \nu & 1 \\ \zero & \nu
  \end{pmatrix} \qquad \text{for some $\nu \in \RR$, for the trace to
    be real.}
\end{equation}

We distinguish several cases.

\subsubsection{$M_H$ diagonalisable with $\mu_1 = \mu_2 = 0$}

To have a deformation at all, it must be that $\theta \neq 0$.  In
that case, $\tilde H := \tfrac1{2\theta}(H - \theta R)$ obeys $[\tilde
H, \B] = i \B$ and $[\tilde H, \P] = 0$.  In summary, the deformation
can be brought to the complex form
\begin{equation}
  \label{eq:zero-zero}
  \boxed{[H,\B] = i \B,}
\end{equation}
or to the real form
\begin{equation}
  \label{eq:zero-zero-real}
  [H, B_a] = -\epsilon_{ab} B_b.
\end{equation}
Although it may not look it, this Lie algebra is isomorphic to the
\textbf{euclidean Newton} algebra, whose real form is typically given
by
\begin{equation}
  [H,B_a] = P_a \qquad\text{and}\qquad [H,P_a] = -B_a.
\end{equation}
Indeed, defining $B'_a = B_a - \epsilon_{ab} P_b$, $P'_a = B_a +
\epsilon_{ab} P_b$ and $H' = -\tfrac12(H + R)$, we see that the
standard Newton algebra brackets imply that the primed generators obey
the Lie brackets in equation~\eqref{eq:zero-zero-real}.

\subsubsection{$M_H$ diagonalisable with $0 \neq \mu_1 \geq \mu_2$}

In this case $\tilde H = \tfrac1{\mu_1} (H + \theta R)$ satisfies
$[\tilde H, \B] = \B$ and $[\tilde H, \P] = (\lambda - 2 i \theta')
\P$, where $\lambda = \mu_2/\mu_1 \leq 1$.  If $\lambda < -1$, we
can let $\tilde H = \frac1{\mu_2} (H + \theta R)$ instead and
exchanging $\B$ and $\P$, so that in any case we can bring the
deformation to the complex form
\begin{equation}
  [H,\B] = \B \qquad\text{and}\qquad [H,\P] = (\lambda + i
    \theta) \P \qquad\text{where $\lambda \in [-1,1]$ and $\theta \in
      \RR$.}
\end{equation}
Moreover we can always assume that $\theta \geq 0$, for if $\theta <0$,
then define $H' = H + 2 \theta R$, $\B'=\P$ and $\P'=\B$ and we
arrive at the the same algebra where $\theta$ has become $-\theta$.
In summary,
\begin{equation}
  \label{eq:zero-nonzero}
  \boxed{[H,\B] = \B \qquad\text{and}\qquad [H,\P] = (\lambda + i
    \theta) \P \qquad\text{where $\lambda \in [-1,1]$ and $\theta\geq 0$,}}
\end{equation}
or in real form
\begin{equation}
  [H,B_a] = B_a \qquad\text{and}\qquad [H,P_a] = \lambda P_a -
  \epsilon_{ab} \theta P_b.
\end{equation}
The case $\lambda = -1$ and $\theta = 0$ is the \textbf{lorentzian
  Newton} algebra.

\subsubsection{$M_H$ nondiagonalisable with nonzero trace}

Since $M_H$ is not diagonalisable, its normal form is given by
\eqref{eq:normal-form-jordan} with $\nu \neq 0$ for nonzero trace.  By
rescaling we can bring the trace to any desired nonzero value, so we may as
well take $\nu = 1$ in \eqref{eq:normal-form-jordan} and arrive at the
Lie brackets in complex form
\begin{equation}
  \label{eq:zero-orbit-nondiag-trace}
  \boxed{[H,\B] = \B  \qquad\text{and}\qquad [H,\P] = \B + \P,}
\end{equation}
and in real form
\begin{equation}
  [H,B_a] = B_a \qquad\text{and}\qquad [H,P_a] = B_a + P_a.
\end{equation}

\subsubsection{$M_H$ nondiagonalisable with zero trace}

In this case, $M_H$ can be conjugated to \eqref{eq:normal-form-jordan}
with $\nu = 0$, leading to the nonzero Lie brackets
\begin{equation}
  \label{eq:galilean}
  \boxed{[H,\P] = \B,}
\end{equation}
which is isomorphic to the \textbf{galilean algebra}.  Usually one
relabels $\B$ and $\P$ and writes the algebra as
\begin{equation}
  [H,B_a] = P_a.
\end{equation}

\subsection{Lightlike orbit}
\label{sec:lightl-orbit}

Here $t_8 = t_{10} = t_{11} = 0$ and $t_9 = 1$.  The obstruction
relations~\eqref{eq:relations} are equivalent to $t_1=t_2=t_4=t_7=0$.
This leaves the following deformation
\begin{equation}
  \begin{split}
    \varphi &= t_3 c_3 + t_5 (c_2 - c_8 + c_9) + t_6 c_4 + c_{10}\\
    &= (t_3 - it_6) \c_2 + t_5 (-i \c_1 + i \c_4 - \c_5) - \c_6 + \text{c.c}
  \end{split}
\end{equation}

In order to bring this to a normal form, it is convenient to use the
subgroup of $G$ which stabilises the vector defining the lightlike
orbit to bring parameters in another $G$-submodule of $\H^2$ to a
simpler form.  In the basis $(\c_{10}, i \cba_8, -i\c_8, \c_6)$, the
lightlike vector labeling this orbit has components $(0,0,0,-2)$.  From
equation~\eqref{eq:4drep-G} we can easily determine that the subgroup
of $G$ which stabilises this vector is given by
\begin{equation}
  \label{eq:lightlike-stab}
  G_{\text{lightlike}} = \left\{ \left(
      \begin{pmatrix}
        a & \zero \\ c & d
      \end{pmatrix}, |a|^2 \right) \in \GL(\CC^2) \times \RR^\times \right\}.
\end{equation}
From equation~\eqref{eq:autos-C2}, we see that a typical element
$(A,\lambda) \in G$ acts on the complex vector subspace spanned by
$(\c_1-\c_4, \c_2,\c_3)$ via
\begin{equation}
  \frac{1}{\lambda \Delta}
  \begin{pmatrix}
    ad + bc & bd & -ac \\ 2cd & d^2 & -c^2 \\ -2ab & -b^2 & a^2
  \end{pmatrix},
\end{equation}
so that a typical element of $G_{\text{lightlike}}$ acts like
\begin{equation}
  \frac{1}{|a|^2}
  \begin{pmatrix}
    1 & \zero & -\frac{c}{d} \\[3pt] 2\frac{c}{a} & \frac{d}{a} &
    -\frac{c^2}{ad} \\[3pt] \zero & \zero & \frac{a}{d}
  \end{pmatrix}.
\end{equation}
The component of the deformation $\varphi$ in that three-dimensional
subspace is parametrised by $(-it_5, t_3-it_6, 0)$, which transforms
under $G_{\text{lightlike}}$ as
\begin{equation}
  \begin{pmatrix}
    -it_5 \\ t_3 - i t_6 \\ \zero
  \end{pmatrix} \mapsto
  \frac1{|a|^2}
  \begin{pmatrix}
    -it_5 \\[3pt] -2i \frac{c}{a} t_5 + \frac{d}{a} (t_3 - i t_6) \\[3pt] \zero
  \end{pmatrix}.
\end{equation}
We distinguish several branches.

\subsubsection{$t_5 =0$ and $t_3 - i t_6 = 0$ branch}

Here $\varphi = -2\c_6 = c_{10}$ and the deformation has additional nonzero
Lie brackets $[B_a,B_b] = \epsilon_{ab} H$.  Rescaling $H$ and using
the complex description, we may write this deformation as
\begin{equation}
  \label{eq:light-1}
  \boxed{[\B,\Bbar] = i H,}
\end{equation}
which is isomorphic (by rescaling $H$ back) to the following real form
\begin{equation}
  [B_a, B_b] = \epsilon_{ab} H.
\end{equation}
\subsubsection{$t_5 =0$ and $t_3 - i t_6 \neq 0$ branch}

Here $t_3 - it_6$ can be brought to $1$, so that the deformation has
additional nonzero Lie brackets
\begin{equation}
  [H,B_a] = P_a \qquad\text{and}\qquad [B_a,B_b] = \epsilon_{ab} H.
\end{equation}
After rescaling $H$ and $P$ and in a complex basis, we arrive at
\begin{equation}
  \label{eq:light-2}
  \boxed{[H,\B] = \P \qquad\text{and}\qquad [\B,\Bbar]= iH.}
\end{equation}

\subsubsection{$t_5 \neq 0$ branch}

Here we can bring $t_3 - it_6$ to zero and $t_5$ to $\pm 1$, resulting
in the Lie brackets
\begin{equation}
  [H,B_a] = \pm \epsilon_{ab} B_b \qquad [H,P_a] = \mp \epsilon_{ab}
  P_b \qquad [B_a,B_b] = \epsilon_{ab} (H \pm R).
\end{equation}
Finally, by redefining $H \pm R$ to be the new $H$ and after rescaling
$B$ and the new $H$, we may bring these brackets to the following
complex form
\begin{equation}
  \label{eq:light-3}
  \boxed{[H,\B] = \pm i \B \qquad\text{and}\qquad [\B,\Bbar] = i H,}
\end{equation}
which is isomorphic to
\begin{equation}
  [H,B_a] = \pm \epsilon_{ab} B_b \qquad\text{and}\qquad [B_a, B_b] =
  \epsilon_{ab} H.
\end{equation}

\subsection{Timelike orbit}
\label{sec:timel-orbit}

Here $t_8= t_{11} =0$, $t_9= 1$ and $t_{10} = -1$.  The obstruction
relations~\eqref{eq:relations} are equivalent to $t_1=t_2 =0$, $t_4 =
- t_3$ and $t_7 = t_6$.  This leaves the following deformation
\begin{equation}
  \begin{split}
    \varphi &= t_3 (c_3 - c_5 + c_{11}) + t_5 (c_2 - c_8 + c_9 - c_{15}) + t_6
    (c_4 + c_6 + c_{12}) + c_{10} + c_{16}\\
    &= (t_3 - it_6)(\c_2 + 2 \c_7) - (t_3 + i t_6) \c_3 + t_5 (-i \c_1 + i \c_4
    - \c_5 + \c_9) - \c_6 - \c_{10} + \text{c.c}
  \end{split}
\end{equation}

In the basis $(\c_{10}, i \cba_8, -i\c_8, \c_6)$, the
timelike vector labeling this orbit has components $(-2,0,0,-2)$.  From
equation~\eqref{eq:4drep-G} we can easily determine that the subgroup
of $G$ which stabilises this vector is given by
\begin{equation}
  \label{eq:timelike-stab}
  G_{\text{timelike}} = \left\{ \left(
      \begin{pmatrix}
        a & b \\ -\gamma \bbar & \gamma \abar
      \end{pmatrix}, |a|^2 + |b|^2 \right) \in \GL(\CC^2) \times
    \RR^\times \middle | \gamma \in \CC\quad |\gamma|=1\right\}.
\end{equation}
From equation~\eqref{eq:autos-C2}, we see that a typical element
$(A,\lambda) \in G$ acts on the complex vector subspace spanned by
$(\c_1-\c_4, \c_2,\c_3)$ via
\begin{equation}
  \frac{1}{\lambda \Delta}
  \begin{pmatrix}
    ad + bc & bd & -ac \\ 2cd & d^2 & -c^2 \\ -2ab & -b^2 & a^2
  \end{pmatrix},
\end{equation}
so that a typical element of $G_{\text{timelike}}$ acts like
\begin{equation}
  \frac{1}{(|a|^2+|b|^2)^2}
  \begin{pmatrix}
    |a|^2-|b|^2 & \abar b & a \bbar \\[3pt] -2\gamma\abar\bbar &
    \gamma\abar^2 & -\gamma \bbar^2\\[3pt] -2\gammabar ab & -\gammabar
    b^2& \gammabar a^2
  \end{pmatrix}.
\end{equation}
The component of the deformation $\varphi$ in that three-dimensional
subspace is parametrised by $(-it_5, t_3-it_6, -t_3 - i t_6)$, which transforms
under $G_{\text{timelike}}$ as
\begin{equation}
  \begin{pmatrix}
    -it_5 \\ t_3 - i t_6 \\ -t_3 - i t_6
  \end{pmatrix} \mapsto
  \frac1{(|a|^2+|b|^2)^2}
  \begin{pmatrix}
    -i (|a|^2-|b|^2) t_5 + \abar b (t_3 -i t_6) - a\bbar (t_3 + i t_6)\\[3pt]
    2i\gamma\abar\bbar t_5 + \gamma\abar^2(t_3-it_6) + \gamma\bbar^2(t_3+it_6)\\[3pt] 
    2i\gammabar a b t_5 - \gammabar b^2(t_3-it_6) - \gammabar a^2(t_3+it_6)
  \end{pmatrix}.
\end{equation}
Acting on $(t_5, t_3, t_6)$ we have
\begin{equation}
  \label{eq:ortho-conf}
  \begin{pmatrix}
    t_5 \\ t_3 \\ t_6
  \end{pmatrix} \mapsto
  \frac1{(|a|^2+|b|^2)^2}
  \begin{pmatrix}
    |a|^2-|b|^2 & -2\Im(\abar b) & 2 \Re(\abar b)\\[3pt]
    -2\Im (\gamma \abar\bbar) & \Re(\gamma(\abar^2+\bbar^2)) & \Im (\gamma(\abar^2-\bbar^2))\\[3pt]
    -2\Re (\gamma \abar\bbar) & -\Im(\gamma(\abar^2+\bbar^2)) & \Re (\gamma(\abar^2-\bbar^2))\\
   \end{pmatrix}
  \begin{pmatrix}
    t_5 \\ t_3 \\ t_6
  \end{pmatrix}.
\end{equation}
The kernel of this representation consists of those matrices with
$|a|=1$, $\gamma=a^2$ and $b=0$, which is a circle subgroup. Therefore
the action is not faithful and only a $4$-dimensional subgroup of
$G_{\text{timelike}}$ acts effectively on $(t_5,t_3,t_6)$. We observe
that the matrix in equation~\eqref{eq:ortho-conf} is conformally
orthogonal; that is, if we let
\begin{equation}
  M := \frac1{(|a|^2+|b|^2)^2}
  \begin{pmatrix}
    |a|^2-|b|^2 & -2\Im(\abar b) & 2 \Re(\abar b)\\[3pt]
    -2\Im (\gamma \abar\bbar) & \Re(\gamma(\abar^2+\bbar^2)) & \Im (\gamma(\abar^2-\bbar^2))\\[3pt]
    -2\Re (\gamma \abar\bbar) & -\Im(\gamma(\abar^2+\bbar^2)) & \Re (\gamma(\abar^2-\bbar^2))\\
   \end{pmatrix},
\end{equation}
then $M^T M = (|a|^2+|b|^2)^{-2} \id$.  We wish to conclude that
$G_{\text{timelike}}$ acts on the three-dimensional space with
coordinates $(t_5, t_3, t_6)$ in such a way that there are two orbits:
the zero vector and all the nonzero vectors.  Since the action is
linear, it is clear that the zero vector is its own orbit, so what we
need to show is that all nonzero vectors lie on the same orbit.  It is
enough to show that the orbit of, say, the vector $(1,0,0)$ under the
orthogonal matrices $(|a|^2+|b|^2) M$, as $a,b,\gamma$ vary, is all of
the unit sphere.  If we write $a = u e^{i \theta}$, $b=v e^{i\psi}$
and $\gamma = e^{i\phi}$, then the image of $(1,0,0)$ under the matrix
$(u^2+v^2) M$ is given by
\begin{equation}
  \left(\frac{u^2-v^2}{u^2+v^2}, \frac{2 u
      v}{u^2+v^2}\cos(\phi-\theta-\psi), \frac{-2u v}{u^2+v^2}
    \sin(\phi-\theta-\psi)\right),
\end{equation}
and if we let $\phi - \theta - \psi = -\vartheta$ and introduce $\rho =-
v/u$ (assuming $u\neq 0$, which is the pole $(-1,0,0)$ corresponding
to the point at infinity), then the above vector becomes
\begin{equation}
  \left(\frac{1-\rho^2}{1+\rho^2}, \frac{2
      \rho}{1+\rho^2}\cos\vartheta, \frac{2\rho}{1+\rho^2}
    \sin\vartheta\right),
\end{equation}
which we recognise as the stereographic projection which parametrises
the unit sphere (minus a pole) in terms of the complex numbers $\rho
e^{i\vartheta}$, up to a relabelling of the coordinates.  Therefore the
action of $G_{\text{timelike}}$ is as claimed and hence acting with
$G_{\text{timelike}}$ we can bring $(t_5, t_3, t_6)$ to one of two
canonical forms: $(0,0,0)$ or $(1,0,0)$, which leads to two different
deformations.

\subsubsection{$(0,0,0)$ normal form}

If $t_3=t_5=t_6=0$, then the deformation is simply $\varphi = -2\c_6 -
2 \c_{10}$, so that rescaling $H$ we can bring the Lie brackets to
\begin{equation}
  \label{eq:time-1}
  \boxed{[\B,\Bbar] = i H \qquad\text{and}\qquad [\P, \Pbar] = i H,}
\end{equation}
which is isomorphic to the following
\begin{equation}
  [B_a, B_b] = \epsilon_{ab} H \qquad\text{and}\qquad [P_a, P_b] =
  \epsilon_{ab} H.
\end{equation}

\subsubsection{$(1,0,0)$ normal form}

On the other hand, if $t_3 = t_6 = 0$ and $t_5 = 1$, the deformation becomes
\begin{equation}
  \begin{split}
    \varphi &= c_2 - c_8 + c_9 - c_{15} + c_{10} + c_{16}\\
    &= -i (\c_1 -\cba_1) + i (\c_4 - \cba_4) - 2\c_5 + 2\c_9 - 2\c_6 - 2\c_{10},
  \end{split}
\end{equation}
which leads to the Lie brackets
\begin{equation}
  [H,\B] = -i \B, \qquad [H,\P] = i\P, \qquad [\B,\Bbar] = -2i (H + R)
  \qquad\text{and}\qquad [\P, \Pbar] = -2i (H - R).
\end{equation}
If we let $H\mapsto -\tfrac12 (R+H)$ and rescale both $\B$ and $\P$ by
a factor of $\tfrac12$, we arrive at the following Lie brackets
\begin{equation}
  \label{eq:time-2}
  \boxed{[H,\B] = i \B, \qquad [\B,\Bbar] = i H \qquad\text{and}\qquad
    [\P,\Pbar] = i (H+R),}
\end{equation}
which is isomorphic to
\begin{equation}
  [H,B_a] = \epsilon_{ab} B_b, \qquad [B_a, B_b] = \epsilon_{ab} H
  \qquad\text{and}\qquad [P_a, P_b] = \epsilon_{ab} (H-R).
\end{equation}

\subsection{Spacelike orbit}
\label{sec:spac-orbit}

In this case $t_8=1$, but $t_9=t_{10} = t_{11} = 0$.  The obstruction
relations \eqref{eq:relations} are equivalent to $t_1 = t_3 = t_4 =
t_5 = 0$.  This leaves the following deformation
\begin{equation}
  \begin{split}
    \varphi &= t_2 (c_1 - c_7 + c_{11}) + t_6 (c_4 + c_9) + t_7 (c_6 +
    c_{15}) + c_{14}\\
    &= t_2 (\c_1 -\c_4 + 2 \c_7) - t_6 (\c_5 + i \c_2) - t_7 (\c_9 + i
    \c_3) - 2 i \c_8 + \text{c.c.}.
  \end{split}
\end{equation}
Relative to the ordered basis $(\c_{10},i\cba_8, -i \c_8, \c_6)$ the
vector labelling this orbit has components $(0,2,2,0)$.
Using~\eqref{eq:4drep-G} we can determine the stabiliser
$G_{\text{spacelike}}$ of this vector and we find that it consists of
the union (not disjoint)
\begin{equation}
  G_{\text{spacelike}} = G' \cup G''
\end{equation}
of the two subsets of $G$ defined by
\begin{equation}
  G' = \left\{\left(
      z \begin{pmatrix}
        1 & i s \\ i t u & u
      \end{pmatrix}, |z|^2 u (1 + s t) \right) \in \GL(\CC^2) \times
    \RR^\times \middle | z \in \CC^\times,~s,t \in \RR,~ st \neq -1,~
    u \in \RR^\times\right\}
\end{equation}
and
\begin{equation}
  G'' = \left\{\left(
      z \begin{pmatrix}
        i s & 1 \\ u & i t u
      \end{pmatrix}, |z|^2 u (1 + s t) \right) \in \GL(\CC^2) \times
    \RR^\times \middle | z \in \CC^\times,~s,t \in \RR,~ st \neq -1,~
    u \in \RR^\times\right\}.
\end{equation}
Using \eqref{eq:autos-C2} we can determine how $G_{\text{spacelike}}$
acts on the three-dimensional real subspace with ordered basis $(\c_1
- \c_4, -i\c_2, -i\c_3)$.  The component of $\varphi$ in this subspace
has coordinates $(t_2, t_6, t_7)$ and we find that under a typical
element of $G' \subset G_{\text{spacelike}}$,
\begin{equation}
  \begin{pmatrix}
    t_2 \\ t_6 \\ t_7
  \end{pmatrix}\mapsto \frac{1}{u^2(1+st)^2|z|^2}
  \begin{pmatrix}
    u(1-st) & us & -ut \\ -2 t u^2 &  u^2 & u^2 t^2 \\ 2s & s^2 & 1
  \end{pmatrix}
    \begin{pmatrix}
    t_2 \\ t_6 \\ t_7
  \end{pmatrix};
\end{equation}
whereas under a typical element of $G''\subset G_{\text{spacelike}}$,
\begin{equation}
  \begin{pmatrix}
    t_2 \\ t_6 \\ t_7
  \end{pmatrix}\mapsto \frac{1}{u^2(1+st)^2|z|^2}
  \begin{pmatrix}
    -u(1-st) & -ut & us \\ 2 t u^2 & u^2 t^2 & u^2 \\ -2s & 1 & s^2
  \end{pmatrix}
    \begin{pmatrix}
    t_2 \\ t_6 \\ t_7
  \end{pmatrix}.
\end{equation}
This action is conformally orthogonal relative to a lorentzian inner
product on this three-dimensional space.  Indeed, if we transform
$(t_2, t_6, t_7)^T$ by either of the two matrices below (rescaled
versions of the matrices in $G'$ and $G''$, respectively),
\begin{equation}
\frac{1}{u(1+st)}
  \begin{pmatrix}
    u(1-st) & us & -ut \\ -2 t u^2 &  u^2 & u^2 t^2 \\ 2s & s^2 & 1
  \end{pmatrix}
  \qquad\qquad
  \frac{1}{u(1+st)}
  \begin{pmatrix}
    -u(1-st) & -ut & us \\ 2 t u^2 & u^2 t^2 & u^2 \\ -2s & 1 & s^2
  \end{pmatrix}
\end{equation}
we find that the indefinite quadratic form $t_2^2 - t_6 t_7$ is
invariant.  Therefore, acting with the matrices in either $G'$ or
$G''$ above, the quadratic form is rescaled by a positive factor
$u^{-2} (1+st)^{-2} |z|^{-4}$.  The determinant of the matrices in
either $G'$ or $G''$ above is given by $u^{-3} (1+st)^{-3}|z|^{-6}$,
which can be either positive or negative.  Consider acting on a vector
with coordinates $(0,t_6, t_7)$. Under $G'$ or $G''$, respectively,
this vector is sent to
\begin{equation}
  \begin{pmatrix}
    t'_2 \\ t'_6 \\ t'_7
  \end{pmatrix} :=\tfrac{1}{u^2(1+st)^2|z|^2}
  \begin{pmatrix}
   u (s t_6 - t t_7)\\ u^2 t_6 + u^2 t^2 t_7 \\ s^2 t_6 + t_7
  \end{pmatrix}
  \quad\text{or}\quad
  \begin{pmatrix}
    t''_2 \\ t''_6 \\ t''_7
  \end{pmatrix} :=\tfrac{1}{u^2(1+st)^2|z|^2}
  \begin{pmatrix}
   u (t t_6 - s t_7)\\ u^2 t^2 t_6 + u^2 t_7 \\ t_6 + s^2 t_7
  \end{pmatrix}
\end{equation}
and therefore if $t_6$ and $t_7$ are positive (resp. negative) so will
be $t'_6$, $t'_7$, $t''_6$ and $t''_7$.  In other words
$G_{\text{spacelike}}$ preserves the time orientation. This means that
the action of $G_{\text{spacelike}}$ on the three-dimensional space
spanned by $(t_2,t_6,t_7)$ defines a homomorphism
$G_{\text{spacelike}} \to \CO(2,1)_+$ whose kernel consists of
elements of the form
\begin{equation}
  \left( \begin{pmatrix}
    z & \zero \\ \zero & z
  \end{pmatrix}, 1 \right)
  \qquad\text{with $|z|=1$}.
\end{equation}
By dimension count, the Lie algebra homomorphism
$\g_{\text{spacelike}} \to \co(2,1)$ is surjective and therefore, by
the Lie correspondence for linear groups (which these clearly are), it
induces a surjective group homomorphism from the identity component of
$G_{\text{spacelike}}$ to that of $\CO(2,1)_+$, which is the proper
orthochronous conformal Lorentz group.  That group, and hence also
$G_{\text{spacelike}}$, acts on the three-dimensional space of
vectors $\t = (t_2,t_6,t_7)$ with the following six orbits, labelled
by the given vector $\t$:
\begin{enumerate}
\item \textbf{zero orbit}, with $\t=(0,0,0)$;
\item \textbf{spacelike orbit}, with $\t=(1,0,0)$;
\item \textbf{past and future lightlike orbits}, with $\t=\pm (0,0,1)$; and
\item \textbf{past and future timelike orbits}, with $\t=\pm (0,1,1)$.
\end{enumerate}

We shall now consider the deformations corresponding to these six
orbits.

\subsubsection{Zero orbit}
\label{sec:zero-orbit-1}

In this case $t_2=t_6=t_7=0$ and hence $\varphi = -2i \c_8 +
\text{c.c}$, which leads (after rescaling $H$) to the Lie brackets
$[\B, \Pbar] = i H$.  However we may simply rotate $\P$ and reabsorb
the $i$ and write this Lie algebra as
\begin{equation}
  \label{eq:space-zero}
  \boxed{[\B,\Pbar] = H,}
\end{equation}
which is isomorphic to
\begin{equation}
  [B_a, P_b] = \delta_{ab} H.
\end{equation}
This is the \textbf{Carroll} algebra.

\subsubsection{Spacelike orbit}
\label{sec:spacelike-orbit}

In this case $t_2=1$ and $t_6 = t_7 = 0$, so that
\begin{equation}
  \varphi = \c_1 - \c_4 + 2 \c_7 -2 i \c_8 + \text{c.c.},
\end{equation}
which leads to the following Lie brackets
\begin{equation}
  \label{eq:space-space}
  \boxed{[H,\B] = \B, \qquad [H,\P] = - \P \qquad\text{and}\qquad
    [\B,\Pbar] = 2 (R - i H),}
\end{equation}
which is isomorphic to
\begin{equation}
  [H,B_a] = B_a, \qquad [H,P_a] = - P_a \qquad\text{and}\qquad [B_a,
  P_b] = \delta_{ab} R + \epsilon_{ab} H.
\end{equation}
This is isomorphic to $\so(3,1)$, which we think of as the de~Sitter
(or hyperbolic) algebra in $2+1$ dimensions.

\subsubsection{Lightlike orbits}
\label{sec:lightlike-orbits}

In this case $t_2= t_6 = 0$ and $t_7 = \tau$, where
$\tau = \pm 1$.  The deformation is
\begin{equation}
  \varphi = - \tau (\c_9 + i \c_3) - 2 i \c_8 + \text{c.c.},
\end{equation}
which leads to the following Lie brackets (after multiplying $\P$ by $-i$)
\begin{equation}
  \label{eq:space-light}
  \boxed{[H,\P] = \tau \B, \qquad [\B,\Pbar] = - 2 H,\qquad\text{and}\qquad [\P,\Pbar] = -2 \tau i R,}
\end{equation}
which is isomorphic to
\begin{equation}
  [H,P_a] = \tau B_a, \qquad [B_a, P_b] = -\delta_{ab} H \qquad\text{and}\qquad [P_a, P_b] = \tau \epsilon_{ab} R.
\end{equation}
These are isomorphic to the euclidean algebra $\e$ for $\tau = 1$ and
the Poincaré algebra $\p$ for $\tau = -1$.

\subsubsection{Timelike orbits}
\label{sec:timelike-orbits}

In this case $t_2=0$ and $t_6 = t_7 = \tau$, where $\tau
= \pm 1$.  The deformation in this case is
\begin{equation}
  \varphi = -\tau (i \c_2 + i\c_3 + \c_5 + \c_9 ) - 2 i \c_8 + \text{c.c.},
\end{equation}
with corresponding Lie brackets given by (after multiplying $\P$ by
$-i$ and $H$ by $\tau$),
\begin{equation}
  \label{eq:space-time}
  \boxed{[H,\B] = - \P,\quad [H,\P] =  \B,\quad  [\B,\Bbar] =
    -2 \tau i R, \quad [\B,\Pbar]= - 2 \tau H
    \quad\text{and}\quad [\P,\Pbar] =  -2 \tau i R,}
\end{equation}
which is isomorphic to
\begin{equation}
  \begin{aligned}[m]
    [H, B_a] &= - P_a\\
    [H, P_a] &= B_a\\
  \end{aligned}
  \qquad\qquad
  \begin{aligned}[m]
    [B_a, B_b] &= \tau \epsilon_{ab} R\\
    [P_a, P_b] &= \tau \epsilon_{ab} R\\
  \end{aligned}
  \qquad\qquad
  [B_a, P_b] = -\tau \delta_{ab} H.
\end{equation}
This is isomorphic to $\so(4)$ for $\tau = 1$ or $\so(2,2)$ for $\tau
= -1$.  We can rescale $\B$ and $\P$ in equation~\eqref{eq:space-time}
in order to eliminate the factors of $2$ from the last three
brackets in the complex form of the algebra, but this reintroduces
some factors of $2$ in the real form of the algebra.

\section{Invariant inner products}
\label{sec:invar-inner-prod}

Recall that a Lie algebra $\k$ is said to be \textbf{metric}, if $\k$
admits a nondegenerate symmetric bilinear form $(-,-): \k \times \k
\to \RR$ satisfying the ``associativity'' condition:
\begin{equation}
  \label{eq:assoc}
  ([x,y],z) = (x,[y,z]) \qquad\forall x,y,z \in \k.
\end{equation}
The Killing form $\kappa(x,y) = \Tr (\ad_x \circ \ad_y)$ is always
associative, but Cartan's semisimplicity criterion says that it is
only nondegenerate for semisimple Lie algebras.  Among the kinematical
Lie algebras found above (and summarised in Table~\ref{tab:summary}),
only $\so(3,1)$, $\so(4)$ and $\so(2,2)$ are semisimple and therefore
metric.  However there are also non-simple kinematical Lie algebras in
the table which are metric.  In this section we investigate the
metricity of the non-simple kinematical Lie algebras in
Table~\ref{tab:summary}.  The results are summarised in the right-most
column of that table.

In determining whether or not a kinematical Lie algebra is metric, it
is more convenient to work with the real form of the Lie algebra.  The
strategy in many cases is simply to exploit the associativity
condition \eqref{eq:assoc} to conclude that no invariant inner product
exists.

To show that the static kinematical Lie algebra \eqref{eq:static-R}
does not admit an invariant inner product, let $X$ be any of $B$ or
$P$.  Then, for any associative symmetric bilinear form,
\begin{equation}
  \epsilon_{ab} (B_b, X_c) = ([R,B_a],X_c) = (R, [B_a,X_c]) = 0.
\end{equation}
Since $B$ can only have nonzero inner product with $B$ or $P$, we find
that $(B_a,-) = 0$ and hence $(-,-)$ is degenerate.  That takes care
of the first row in Table~\ref{tab:summary}.  The next five rows in
Table~\ref{tab:summary} describe Lie algebras where the ideal spanned
by $B$ and $P$ is abelian.  The exact same argument as for the static
kinematical Lie algebra shows that any associative symmetric bilinear
form is degenerate.  Finally, a similar argument shows that neither do
the Lie algebras \eqref{eq:light-1}, \eqref{eq:light-2} and
\eqref{eq:light-3}, corresponding to the last three rows in Table
\ref{tab:summary}, admit invariant inner products.  Indeed, if $(-,-)$
is an associative symmetric bilinear form, then if $X$ stands for
either $B$ or $P$, we have
\begin{equation}
  \epsilon_{ab} (P_b, X_c) = ([R,P_a],X_c) = (R, [P_a,X_c]) = 0,
\end{equation}
so that $(P_a,-) = 0$.

In this dimension, the Carroll, Poincaré and euclidean algebras are
metric.  The Carroll algebra \eqref{eq:space-zero} admits a
two-parameter family of invariant inner products:
\begin{equation}
  \label{eq:carroll-ip}
  (B_a, P_b) = \epsilon_{ab} \lambda \qquad  (H,R) = \lambda \qquad (R,R)
  = \mu \qquad \forall \lambda,\mu \in \RR,\quad \lambda \neq 0.
\end{equation}
The euclidean algebra \eqref{eq:space-light} ($\tau = 1$) also admits
a two-parameter family of invariant inner products:
\begin{equation}
  \label{eq:euclidean-ip}
  (B_a, P_b) = \epsilon_{ab} \lambda \qquad (P_a, P_b) = \delta_{ab}
  \mu \qquad  (H,R) = \lambda \qquad (R,R) = \mu \qquad \forall
  \lambda,\mu \in \RR,\quad \lambda \neq 0,
\end{equation}
and so does the Poincaré algebra \eqref{eq:space-light} ($\tau = -1$):
\begin{equation}
  \label{eq:poincare-ip}
  (B_a, P_b) = \epsilon_{ab} \lambda \qquad (P_a, P_b) = -\delta_{ab}
  \mu \qquad  (H,R) = \lambda \qquad (R,R) = \mu \qquad \forall
  \lambda,\mu \in \RR,\quad \lambda \neq 0.
\end{equation}

Finally, the kinematical Lie algebras \eqref{eq:time-1} and
\eqref{eq:time-2}, which are unique to this dimension, are also
metric.  Indeed, the former algebra has a two-parameter family of
invariant inner products given by
\begin{equation}
  \label{eq:time-1-ip}
  (B_a, B_b) = \delta_{ab} \lambda \qquad (P_a, P_b) = \delta_{ab}
  \lambda \qquad  (H,R) = \lambda \qquad (R,R) = \mu \qquad \forall
  \lambda,\mu \in \RR,\quad \lambda \neq 0,
\end{equation}
and does the latter algebra, whose inner product is given by
\begin{equation}
  \label{eq:time-2-ip}
  (B_a, B_b) = \delta_{ab} \lambda \qquad (P_a, P_b) = \delta_{ab}
  (\lambda-\mu) \qquad  (H,R) = \lambda \qquad (H,H) = \lambda \qquad (R,R) = \mu,
\end{equation}
for all $\lambda,\mu \in \RR$ with $\lambda \neq 0$ and $\lambda \neq \mu$.

\section{Summary}
\label{sec:summary}

We have classified all kinematical real Lie algebras in dimension
$2+1$ (up to Lie algebra isomorphism) by classifying the deformations
of the static kinematical Lie algebra, using the approach advocated in
\cite{JMFKinematical3D} and used in \cite{JMFKinematicalHD} to
classify all kinematical Lie algebras in dimension $D+1$ for $D \geq
4$.  Since for $D<2$ the kinematical condition on a Lie algebra is
vacuous, except for specifying the dimension, the results of this
paper complete the classification of kinematical Lie algebras in any
dimension.  It should perhaps be remarked that in physical/geometrical
applications, it is desirable to refine this classification and
distinguish kinematical Lie algebras which, although isomorphic as Lie
algebras, act differently on the ($2+1$)-dimensional spacetime.  This
finer classification is the subject of a forthcoming paper containing the
classification of spacetimes for kinematical Lie algebras in all
dimensions.

Table~\ref{tab:summary} displays the classification for $D=2$. All Lie
brackets are written in the complex form and share the brackets in
equation~\eqref{eq:static-C}, which are not written explicitly. We
also have the Lie brackets obtained from the ones shown via complex
conjugation, but we do not write them explicitly either. Thus the
table contains the minimal data necessary to reconstruct the Lie
algebras.  In some cases, we have relabelled $\B$ and $\P$ in order to
make the description more uniform.  The Lie algebras below the line
are unique to $D=2$, whereas those above the line are $D=2$ versions
of kinematical Lie algebras which occur also for any $D > 2$.  In
$D=3$ there are also some kinematical Lie algebras which have no
analogue in any other dimension: there, due to the existence of the
rotationally invariant vector product in $\RR^3$, whereas the
kinematical Lie algebras unique to $D=2$ owe their existence to the
rotationally invariant symplectic structure on $\RR^2$.

\begin{table}[h!]\tiny
  \centering
  \setlength{\extrarowheight}{2pt}  
  \caption{Kinematical Lie algebras in $2+1$ dimensions (complex form)}
  \label{tab:summary}
  \begin{tabular}{l|*{5}{>{$}l<{$}}|l|c}
    \multicolumn{1}{c|}{Eq.} & \multicolumn{5}{c|}{Nonzero Lie brackets} & \multicolumn{1}{c|}{Comments} & \multicolumn{1}{c}{Metric?}\\\hline
    \ref{eq:static-C} & & & & & & static\\
    \ref{eq:galilean} & [H,\B] = \P & & & & & galilean & \\
    \ref{eq:zero-orbit-nondiag-trace} & [H,\B] = \B & [H,\P] = \B + \P & & & &  & \\
    \ref{eq:zero-nonzero} & [H,\B] = \B & [H,\P] = \P & & & & & \\
    \ref{eq:zero-nonzero} & [H,\B] = \B & [H,\P] = (1 + i \theta) \P & & & & $\theta>0$ & \\
    \ref{eq:zero-nonzero} & [H,\B] = \B & [H,\P] = \lambda \P & & & & $\lambda \in (-1,1)$ & \\
    \ref{eq:zero-nonzero} & [H,\B] = \B & [H,\P] = -\P & & & &lorentzian Newton & \\
    \ref{eq:zero-zero} & [H,\B] = i \B & & & & & euclidean Newton & \\
    \ref{eq:space-zero} & & & & [\B,\Pbar] = 2 H & & Carroll & \checkmark \eqref{eq:carroll-ip}\\
    \ref{eq:space-light} & & [H,\P] = -\B & & [\B,\Pbar] = 2 H & [\P,\Pbar] = -2 i R & $\e$ & \checkmark \eqref{eq:euclidean-ip}\\
    \ref{eq:space-light} & & [H,\P] = \B & & [\B,\Pbar] = 2 H & [\P,\Pbar] = 2 i R & $\p$ & \checkmark \eqref{eq:poincare-ip}\\
    \ref{eq:space-space} & [H,\B] = \B & [H,\P] = -\P & & [\B,\Pbar] = 2(H + i R) & & $\so(3,1)$ & \checkmark \\
    \ref{eq:space-time} & [H,\B] = \P & [H,\P] =  -\B & [\B,\Bbar] =  - 2 i R & [\B,\Pbar]= 2 H & [\P,\Pbar] =  - 2 i R & $\so(4)$ & \checkmark \\
    \ref{eq:space-time} & [H,\B] = -\P & [H,\P] =  \B & [\B,\Bbar] =   2 i R & [\B,\Pbar]=  2 H & [\P,\Pbar] =   2 i R & $\so(2,2)$ & \checkmark \\\hline
    \ref{eq:zero-nonzero} & [H,\B] = \B & [H,\P] = (\lambda + i \theta) \P & & & & $\lambda \in (-1,1)$ and $\theta>0$ & \\
    \ref{eq:time-1} & & & [\B, \Bbar]=i H & & [\P, \Pbar] = i H & & \checkmark \eqref{eq:time-1-ip} \\
    \ref{eq:time-2} & [H,\B] = i \B & & [\B,\Bbar] = i H & & [\P,\Pbar] = i (H+R) &  & \checkmark \eqref{eq:time-2-ip}\\
    \ref{eq:light-1} & & & [\B,\Bbar] = i H & & &  & \\
    \ref{eq:light-2} & [H,\B] = \P & & [\B,\Bbar] = i H & & &  & \\
    \ref{eq:light-3} & [H,\B] = \pm i \B & & [\B,\Bbar] = i H & & &  & \\
  \end{tabular}
\end{table}

The first six lines consist of Lie algebras which are the semidirect
product of the abelian subalgebra generated by $H$ and $R$ and a
four-dimensional real representation (real and imaginary parts of a
two-dimensional complex representation spanned by $\B$ and $\P$),
where representation where $R$ acts as multiplication by $-i$ and $H$,
which commutes with $R$ therefore acts complex linearly. This means
that the action of $H$ (relative to the basis $\B$ and $\P$) is
characterised by a $2\times 2$ complex matrix. However not every such
matrix gives rise to different (i.e., non-isomorphic) semidirect
products. We can change basis $(\B,\P) \mapsto (\B', \P')$, which is
the same as conjugating the matrix of $H$ in $\GL(2,\CC)$, but we can
also modify $H$ itself by affine transformations of the form
$H \mapsto \lambda H + \mu R$, where $\lambda,\mu \in \RR$ and
$\lambda \neq 0$. The first six lines in the table correspond
precisely to the isomorphism classes of such semidirect products.

\section*{Acknowledgments}
\label{sec:acknowledgments}

The work of JMF is partially supported by the grant ST/L000458/1
``Particle Theory at the Higgs Centre'' from the UK Science and
Technology Facilities Council.

\appendix

\section{Enumerations of the deformation complex}
\label{app:enum-complex}

In this appendix we enumerate the first few graded subspaces of the
deformation complexes $C^\bullet$ and $C^\bullet_\CC$, which we will
refer to informally as the real and complex deformations complexes.

\subsection{Enumeration of the real deformation complex}
\label{sec:enum-real-deform}

We shall now enumerate bases for $C^p$, $p=0,1,2,3$, and the dimension
count in Section~\ref{sec:def-comp} will ensure that we have not left
out any basis elements.  $C^0$ is spanned by $R$ and $H$.  Bases for $C^1$, $C^2$
and $C^3$ are tabulated below in abbreviated form, where we
distinguish between $\beta\pi = \beta_a\pi_a$ and $\epsilon\beta\pi =
\epsilon_{ab}\beta_a \pi_b$, et cetera.  In particular, we can now
have $\epsilon\beta\beta =  \epsilon_{ab}\beta_a\beta_b$.  Similarly,
we must distinguish between $\beta B = \beta_a B_a$ and $\epsilon\beta
B = \epsilon_{ab}\beta_a B_b$.  Notice however that for any $X \in
\g$, $\epsilon \beta\pi \beta X$ and $\epsilon\beta\beta \pi X$ are
collinear, et cetera.  Similarly, any terms with two $\epsilon$ can be
rewritten with no $\epsilon$'s using the identity
$\epsilon_{ab}\epsilon_{cd} = \delta_{ac}\delta_{bd} -
\delta_{bc}\delta_{ad}$.

\begin{table}[h!]
  \centering
  \caption{Basis for $C^1(\h;\g)^{\r}$}
  \label{tab:basis-C1}
  \begin{tabular}{*{10}{>{$}c<{$}}}
    a_1& a_2 & a_3 & a_4 & a_5 & a_6 & a_7 & a_8 & a_9 & a_{10}  \\\hline
    \eta R & \eta H & \beta B & \epsilon\beta B & \beta P & \epsilon\beta P & \pi B & \epsilon\pi B& \pi P & \epsilon\pi P\\
  \end{tabular}
\end{table}

\begin{table}[h!]
  \centering
  \caption{Basis for $C^2(\h;\g)^{\r}$}
  \label{tab:basis-C2}
  \begin{tabular}{*{8}{>{$}c<{$}}}
    c_1 & c_2 & c_3 & c_4 & c_5 & c_6 & c_7 & c_8 \\\hline
    \eta\beta B & \eta \epsilon\beta B & \eta\beta P & \eta \epsilon\beta P & \eta\pi B & \eta \epsilon\pi B & \eta\pi P & \eta \epsilon\pi P \\[10pt]
    c_9 & c_{10} & c_{11} & c_{12} & c_{13} & c_{14} & c_{15} & c_{16} \\\hline
    \tfrac12 \epsilon\beta\beta R & \tfrac12 \epsilon\beta\beta H & \beta \pi R & \epsilon\beta\pi R & \beta\pi H & \epsilon \beta \pi H & \tfrac12 \epsilon\pi\pi R & \tfrac12 \epsilon \pi \pi H\\
  \end{tabular}
\end{table}

\begin{table}[h!]
  \centering
  \caption{Basis for $C^3(\h;\g)^{\r}$}
  \label{tab:basis-C3}
  \begin{tabular}{*{8}{>{$}c<{$}}}
    b_1 & b_2 & b_3 & b_4 & b_5 & b_6 & b_7 & b_8 \\\hline
    \eta\epsilon\beta\beta R & \eta \epsilon \beta\beta H & \eta \epsilon\pi\pi R & \eta\epsilon \pi\pi H & \eta\beta\pi R & \eta\epsilon\beta\pi R & \eta \beta \pi H & \eta\epsilon\beta\pi H \\[10pt]
    b_9 & b_{10} & b_{11}  & b_{12} &  b_{13} &  b_{14} &  b_{15} & b_{16} \\\hline 
    \epsilon\beta\beta\pi B & \epsilon\beta\beta \pi P & \epsilon\pi\pi\beta B & \epsilon\pi\pi\beta P & \beta \pi \beta B & \beta \pi \beta P & \beta \pi \pi B & \beta\pi \pi P \\
  \end{tabular}
\end{table}

Finally we work out the Nijenhuis--Richardson bracket $C^2 \times C^2
\to C^3$.  Table~\ref{tab:NR-dot} displays the multiplication table
for $\bullet: C^2 \times C^2 \to C^3$ from where we obtain the
Nijenhuis--Richardson bracket by symmetrisation:
\begin{equation}\small
  \label{eq:NR-bracket-C2}
  \begin{aligned}[m]
    [\![c_1,c_9]\!]&= b_{1}\\
    [\![c_1,c_{10}]\!]&= b_{2}\\
    [\![c_1,c_{11}]\!]&= b_{5}\\
    [\![c_1,c_{12}]\!]&= b_{6}\\
    [\![c_1,c_{13}]\!]&= b_{7}+b_{13}\\
    [\![c_1,c_{14}]\!]&= b_{8}-\tfrac12 b_{9}\\
    [\![c_1,c_{16}]\!]&= \tfrac12 b_{11}\\
    [\![c_2,c_{11}]\!]&= b_{6}\\
    [\![c_2,c_{12}]\!]&= -b_{5}\\
    [\![c_2,c_{13}]\!]&= b_{8}+ \tfrac12 b_{9}\\
    [\![c_2,c_{14}]\!]&= -b_{7}+b_{13}\\
  \end{aligned}
  \qquad\qquad
  \begin{aligned}[m]
    [\![c_2,c_{16}]\!]&= b_{15}\\
    [\![c_3,c_{12}]\!]&= b_{1}\\
    [\![c_3,c_{13}]\!]&= b_{14}\\
    [\![c_3,c_{14}]\!]&= b_{2}- \tfrac12 b_{10}\\
    [\![c_3,c_{15}]\!]&= b_{6}\\
    [\![c_3,c_{16}]\!]&= b_{8}+ \tfrac12 b_{12}\\
    [\![c_4,c_{11}]\!]&= -b_{1}\\
    [\![c_4,c_{13}]\!]&= -b_{2}+ \tfrac12 b_{10}\\
    [\![c_4,c_{14}]\!]&= b_{14}\\
    [\![c_4,c_{15}]\!]&= -b_{5}\\
    [\![c_4,c_{16}]\!]&= -b_{7}+b_{16}\\
  \end{aligned}
  \qquad\qquad
  \begin{aligned}[m]
    [\![c_5,c_9]\!]&= b_{6}\\
    [\![c_5,c_{10}]\!]&= b_{8}+ \tfrac12 b_{9}\\
    [\![c_5,c_{12}]\!]&= b_{3}\\
    [\![c_5,c_{13}]\!]&= b_{15}\\
    [\![c_5,c_{14}]\!]&= b_{4}- \tfrac12 b_{11}\\
    [\![c_6,c_9]\!]&= b_{5}\\
    [\![c_6,c_{10}]\!]&= b_{7}-b_{13}\\
    [\![c_6,c_{11}]\!]&= b_{3}\\
    [\![c_6,c_{13}]\!]&= b_{4}- \tfrac12 b_{11}\\
    [\![c_6,c_{14}]\!]&= -b_{15}\\
    [\![c_7,c_{10}]\!]&=  \tfrac12 b_{10}\\
  \end{aligned}
  \qquad\qquad
  \begin{aligned}[m]
    [\![c_7,c_{11}]\!]&= b_{5}\\
    [\![c_7,c_{12}]\!]&= b_{6}\\
    [\![c_7,c_{13}]\!]&= b_{7}+b_{16}\\
    [\![c_7,c_{14}]\!]&= b_{8}- \tfrac12 b_{12}\\
    [\![c_7,c_{15}]\!]&= b_{3}\\
    [\![c_7,c_{16}]\!]&= b_{4}\\
    [\![c_8,c_{10}]\!]&= -b_{14}\\
    [\![c_8,c_{11}]\!]&= -b_{6}\\
    [\![c_8,c_{12}]\!]&= b_{5}\\
    [\![c_8,c_{13}]\!]&= -b_{8}- \tfrac12 b_{12}\\
    [\![c_8,c_{14}]\!]&= b_{7}-b_{16}\\
  \end{aligned}
\end{equation}

\begin{table}[h!]
  \setlength{\extrarowheight}{4pt}
  \setlength{\tabcolsep}{2pt}
  \centering
  \caption{Nijenhuis--Richardson $\bullet : C^2 \times C^2 \to C^3$}
  \label{tab:NR-dot}
  \begin{tabular}{>{$}c<{$}|*{16}{>{$}c<{$}}}
    \bullet & c_1 & c_2 & c_3 & c_4 & c_5 & c_6 & c_7 & c_8 & c_9 & c_{10} & c_{11} & c_{12} & c_{13} & c_{14} & c_{15} & c_{16} \\\hline
    c_{1} & \zero & \zero & \zero & \zero & \zero & \zero & \zero & \zero & b_1 & b_2 & b_5 & b_6 & b_7 & b_8 & \zero & \zero\\
    c_{2} & \zero & \zero & \zero & \zero & \zero & \zero & \zero & \zero & \zero & \zero & b_6 & -b_5 & b_8 & -b_7 & \zero & \zero\\
    c_{3} & \zero & \zero & \zero & \zero & \zero & \zero & \zero & \zero & \zero & \zero & \zero & b_1 & \zero & b_2 & b_6 & b_8 \\
    c_{4} & \zero & \zero & \zero & \zero & \zero & \zero & \zero & \zero & \zero & \zero & -b_1 & \zero & -b_2 & \zero & -b_5 & -b_7 \\
    c_{5} & \zero & \zero & \zero & \zero & \zero & \zero & \zero & \zero & b_6 & b_8 & \zero & b_3 & \zero & b_4 & \zero & \zero\\
    c_{6} & \zero & \zero & \zero & \zero & \zero & \zero & \zero & \zero & b_5 & b_7 & b_3 & \zero & b_4 & \zero & \zero & \zero\\
    c_{7} & \zero & \zero & \zero & \zero & \zero & \zero & \zero & \zero & \zero & \zero & b_5 & b_6 & b_7 & b_8 & b_3 & b_4 \\
    c_{8} & \zero & \zero & \zero & \zero & \zero & \zero & \zero & \zero & \zero & \zero & -b_6 & b_5 & -b_8 & b_7 &\zero & \zero \\
    c_{9} & \zero & \zero & \zero & \zero & \zero & \zero & \zero & \zero & \zero & \zero & \zero & \zero & \zero & \zero & \zero & \zero\\
    c_{10} & \zero & \zero & \zero & \zero & \frac12 b_9 & -b_{13} & \frac12 b_{10} & -b_{14} & \zero & \zero & \zero & \zero & \zero & \zero & \zero & \zero \\    
    c_{11} & \zero & \zero & \zero & \zero & \zero & \zero & \zero & \zero & \zero & \zero & \zero & \zero & \zero & \zero & \zero & \zero\\
    c_{12} & \zero & \zero & \zero & \zero & \zero & \zero & \zero & \zero & \zero & \zero & \zero & \zero & \zero & \zero & \zero & \zero\\
    c_{13} & b_{13} & \frac12 b_9 & b_{14} & \frac12 b_{10} & b_{15} & -\frac12 b_{11} & b_{16} & -\frac12 b_{12} & \zero & \zero & \zero & \zero & \zero & \zero & \zero & \zero\\
    c_{14} &  -\frac12 b_9 & b_{13} & -\frac12 b_{10} & b_{14} & -\frac12 b_{11} & -b_{15} & -\frac12 b_{12} & -b_{16} & \zero & \zero & \zero & \zero & \zero & \zero & \zero & \zero\\
    c_{15} & \zero & \zero & \zero & \zero & \zero & \zero & \zero & \zero & \zero & \zero & \zero & \zero & \zero & \zero & \zero & \zero\\
    c_{16} & \frac12 b_{11} & b_{15} & \frac12 b_{12} & b_{16}& \zero & \zero & \zero & \zero & \zero & \zero & \zero & \zero & \zero & \zero & \zero & \zero\\
  \end{tabular}
\end{table}

\subsection{Enumeration of the complex deformation complex}
\label{sec:enum-compl-deform}

We shall now enumerate bases for $C_\CC^p$, $p=0,1,2,3$.  $C_\CC^0$ is
spanned by $R$ and $H$.  A basis for $C^1_\CC$ is given by $\a_1$,
$\a_2$, $\a_3$, $\aba_3$, $\a_4$, $\aba_4$, $\a_5$, $\aba_5$, $\a_6$
and $\aba_6$, where the $\a_i$ are defined in
Table~\ref{tab:basis-C1-C}.  A basis for $C_\CC^2$ is given by $\c_1$,
$\cba_1$, $\c_2$, $\cba_2$, $\c_3$, $\cba_3$, $\c_4$, $\cba_4$,
$\c_5$, $\c_6$, $\c_7$, $\cba_7$, $\c_8$, $\cba_8$, $\c_9$ and
$\c_{10}$, where the $\c_i$ are defined in Table~\ref{tab:basis-C2-C}.
Finally, as basis for $C_\CC^3$ is given by $\b_1$, $\b_2$, $\b_3$,
$\bba_3$, $\b_4$, $\bba_4$, $\b_5$, $\b_6$, $\b_7$, $\bba_7$, $\b_8$,
$\bba_8$, $\b_9$, $\bba_9$, $\b_{10}$ and $\bba_{10}$, where the
$\b_i$ are defined in Table~\ref{tab:basis-C3-C}.  The complex
conjugates of the basis elements are the naive ones, e.g.,
$\cba_1 = \eta\betabar\Bbar$.

\begin{table}[h!]
  \setlength{\extrarowheight}{3pt}
  \centering
  \caption{Basis for $C^1(\h_\CC;\g_\CC)^{\r}$}
  \label{tab:basis-C1-C}
  \begin{tabular}{*{6}{>{$}c<{$}}}
    \a_1& \a_2 & \a_3 & \a_4 & \a_5 & \a_6 \\\hline
    \eta R & \eta H & \bbeta \B & \bbeta \P & \bpi \B & \bpi \P
  \end{tabular}
\end{table}

\begin{table}[h!]
  \setlength{\extrarowheight}{3pt}
  \centering
  \caption{Basis for $C^2(\h_\CC;\g_\CC)^{\r}$}
  \label{tab:basis-C2-C}
  \begin{tabular}{*{10}{>{$}c<{$}}}
    \c_1 & \c_2 & \c_3 & \c_4 & \c_5 & \c_6 & \c_7 & \c_8 & \c_9 & \c_{10}\\\hline
    \eta\bbeta \B & \eta\bbeta \P & \eta\bpi \B & \eta\bpi \P & i \bbeta\betabar R & i\bbeta\betabar H & \bbeta\pibar R & \bbeta\pibar H & i\bpi\pibar R & i \bpi\pibar H
  \end{tabular}
\end{table}

\begin{table}[h!]
  \setlength{\extrarowheight}{3pt}
  \centering
  \caption{Basis for $C^3(\h_\CC;\g_\CC)^{\r}$}
  \label{tab:basis-C3-C}
  \begin{tabular}{*{10}{>{$}c<{$}}}
    \b_1 & \b_2 & \b_3 & \b_4 & \b_5 & \b_6 & \b_7 & \b_8 & \b_9 & \b_{10} \\\hline 
    i \eta\bbeta\betabar R & i\eta\bbeta\betabar H & \eta\bbeta\pibar R & \eta\bbeta\pibar H & i\eta\bpi\pibar R &  i\eta\bpi\pibar H & i\bbeta\betabar\bpi\B & i\bbeta\betabar\bpi\P & i\bbeta\bpi\pibar\B & i\bbeta\bpi\pibar\P
  \end{tabular}
\end{table}

Finally we work out the Nijenhuis--Richardson bracket $C_\CC^2 \times C_\CC^2
\to C_\CC^3$.  Table~\ref{tab:NR-dot-C} displays the multiplication table
for $\bullet: C_\CC^2 \times C_\CC^2 \to C_\CC^3$ from where we obtain the
Nijenhuis--Richardson bracket by symmetrisation.

\begin{table}[h!]
  \setlength{\extrarowheight}{4pt}
  \setlength{\tabcolsep}{2pt}
  \centering
  \caption{Nijenhuis--Richardson $\bullet : C_\CC^2 \times C_\CC^2 \to C_\CC^3$}
  \label{tab:NR-dot-C}
  \begin{tabular}{>{$}c<{$}|*{16}{>{$}c<{$}}}
    \bullet & \c_1 & \cba_1 & \c_2 & \cba_2 & \c_3 & \cba_3 & \c_4 & \cba_4 & \c_5 & \c_6 & \c_7 & \cba_7 & \c_8 & \cba_8 & \c_9 & \c_{10} \\\hline
    \c_1 & \zero & \zero & \zero & \zero & \zero & \zero & \zero & \zero & \b_1 & \b_2 & \b_3 & \zero & \b_4 & \zero & \zero & \zero\\
    \cba_1 & \zero & \zero & \zero & \zero & \zero & \zero & \zero & \zero & \b_1 & \b_2 & \zero & \bba_3 & \zero & \bba_4 & \zero & \zero\\
    \c_2 & \zero & \zero & \zero & \zero & \zero & \zero & \zero & \zero & \zero & \zero & \zero & i \b_1 & \zero & i \b_2 & i \b_3 & i \b_4 \\
    \cba_2 & \zero & \zero & \zero & \zero & \zero & \zero & \zero & \zero & \zero & \zero & -i \b_1 & \zero & -i \b_2 & \zero & -i \bba_3 & - i\bba_4 \\
    \c_3 & \zero & \zero & \zero & \zero & \zero & \zero & \zero & \zero & -i\bba_3 & -i\bba_4 & -i\b_5 & \zero & -i\b_6 & \zero & \zero & \zero\\
    \cba_3 & \zero & \zero & \zero & \zero & \zero & \zero & \zero & \zero & i\b_3 & i\b_4 & \zero & i\b_5 & \zero & i\b_6 & \zero & \zero\\
    \c_4 & \zero & \zero & \zero & \zero & \zero & \zero & \zero & \zero & \zero & \zero & \zero & \bba_3 & \zero & \bba_4 & \b_5 & \b_6 \\
    \cba_4 & \zero & \zero & \zero & \zero & \zero & \zero & \zero & \zero & \zero & \zero & \b_3 & \zero & \b_4 & \zero & \b_5 & \b_6 \\
    \c_5 & \zero & \zero & \zero & \zero & \zero & \zero & \zero & \zero & \zero & \zero & \zero & \zero & \zero & \zero & \zero & \zero\\
    \c_6 & \zero & \zero & \zero & \zero & \b_7 & \bba_7 & \b_8 & \bba_8 & \zero & \zero & \zero & \zero & \zero & \zero & \zero & \zero \\    
    \c_7 & \zero & \zero & \zero & \zero & \zero & \zero & \zero & \zero & \zero & \zero & \zero & \zero & \zero & \zero & \zero & \zero\\
    \cba_7 & \zero & \zero & \zero & \zero & \zero & \zero & \zero & \zero & \zero & \zero & \zero & \zero & \zero & \zero & \zero & \zero\\
    \c_8 & \zero & i\bba_7 & \zero & i\bba_8 & i \b_9 & \zero & i\b_{10} & \zero & \zero & \zero & \zero & \zero & \zero & \zero & \zero & \zero\\
    \cba_8 & -i \b_7 & \zero & -i \b_8 & \zero & \zero & -i\bba_9 & \zero  & -i\bba_{10} & \zero & \zero & \zero & \zero & \zero & \zero & \zero & \zero\\
    \c_9 & \zero & \zero & \zero & \zero & \zero & \zero & \zero & \zero & \zero & \zero & \zero & \zero & \zero & \zero & \zero & \zero\\
    \c_{10} & \b_9 & \bba_9 & \b_{10} & \bba_{10}& \zero & \zero & \zero & \zero & \zero & \zero & \zero & \zero & \zero & \zero & \zero & \zero\\
  \end{tabular}
\end{table}

The nonzero Nijenhuis--Richardson brackets are the following:
\begin{equation}\small
  \label{eq:NR-brackets-C2-C}
  \begin{aligned}[m]
    [\![\c_1,\c_5]\!] &= \b_1\\
    [\![\c_1,\c_6]\!] &= \b_2\\
    [\![\c_1,\c_7]\!] &= \b_3\\
    [\![\c_1,\c_8]\!] &= \b_4\\
    [\![\c_1,\cba_8]\!] &= -i\b_7\\
  \end{aligned}
  \qquad\qquad
  \begin{aligned}[m]
    [\![\c_2,\cba_7]\!] &= i \b_1\\
    [\![\c_2,\cba_8]\!] &= i(\b_2 - \b_8)\\
    [\![\c_2,\c_9]\!] &= i\b_3\\
    [\![\c_2,\c_{10}]\!] &= i\b_4 + \b_{10}\\
  \end{aligned}
  \qquad\qquad
  \begin{aligned}[m]
    [\![\c_3,\c_5]\!] &= -i\bba_3\\
    [\![\c_3,\c_6]\!] &= -i\bba_4 + \b_7\\
    [\![\c_3,\c_7]\!] &= -i\b_5\\
    [\![\c_3,\c_8]\!] &= -i(\b_6-\b_9)\\
  \end{aligned}
  \qquad\qquad
  \begin{aligned}[m]
    [\![\c_4,\c_6]\!] &= \b_8\\
    [\![\c_4,\c_7]\!] &= \bba_3\\
    [\![\c_4,\c_8]\!] &= i\b_{10}\\
    [\![\c_4,\cba_8]\!] &= \bba_4\\
    [\![\c_4,\cba_7]\!] &= \b_5\\
    [\![\c_4,\cba_7]\!] &= \b_6
  \end{aligned}
\end{equation}
and their complex conjugates, which we do not list explicitly.  For
example, $[\![\cba_1,\c_5]\!]= \b_1$, et cetera, using that
$[\![\bar\lambda,\bar\mu]\!] = \overline{[\![\lambda,\mu]\!]}$.

\subsection{Dictionary between the two enumerations}
\label{sec:dict-betw-two}

For ease of translation between the complex and real enumerations, we
provide the following dictionary for the first two spaces of
cochains.  For $C^1$ we have
\begin{equation}
  \label{eq:dictionary-C1}
  \begin{aligned}[m]
    a_1 &= \a_1\\
    a_2 &= \a_2\\
    a_3 &= \a_3 + \aba_3\\
    a_4 &= -i(\a_3 - \aba_3)\\
    a_5 &= \a_4 + \aba_4\\
    a_6 &= -i(\a_4 - \aba_4)\\
    a_7 &= \a_5 + \aba_5\\
    a_8 &= -i(\a_5 - \aba_5)\\
    a_9 &= \a_6 + \aba_6\\
    a_{10} &= -i(\a_6 - \aba_6)\\
  \end{aligned}
  \qquad\qquad
  \begin{aligned}[m]
    \a_1 &=  a_1 \\
    \a_2 &=  a_2 \\
    \a_3 &= \tfrac12 (a_3 + i a_4)\\
    \a_4 &= \tfrac12 (a_5 + i a_6)\\
    \a_5 &= \tfrac12 (a_7 + i a_8)\\
    \a_6 &= \tfrac12 (a_9 + i a_{10})\\
  \end{aligned}
\end{equation}
and for $C^2$ we have
\begin{equation}
  \label{eq:dictionary-C2}
  \begin{aligned}[m]
    c_1 &= \c_1 + \cba_1\\
    c_2 &= -i (\c_1 - \cba_1)\\
    c_3 &= \c_2 + \cba_2\\
    c_4 &= -i (\c_2 - \cba_2)\\
    c_5 &= \c_3 + \cba_3\\
    c_6 &= -i (\c_3 - \cba_3)\\
    c_7 &= \c_4 + \cba_4\\
    c_8 &= -i (\c_4 - \cba_4)\\
    c_9 &= -2\c_5 \\
    c_{10} &= -2\c_6\\
    c_{11} &= 2 (\c_7 + \cba_7)\\
    c_{12} &= -2i (\c_7 - \cba_7)\\
    c_{13} &= 2 (\c_8 + \cba_8)\\
    c_{14} &= -2i (\c_8 - \cba_8)\\
    c_{15} &= -2\c_9\\
    c_{16} &= -2\c_{10}\\
  \end{aligned}
  \qquad\qquad
  \begin{aligned}[m]
    \c_1 &= \tfrac12 (c_1 + i c_2)\\
    \c_2 &= \tfrac12 (c_3 + i c_4)\\
    \c_3 &= \tfrac12 (c_5 + i c_6)\\
    \c_4 &= \tfrac12 (c_7 + i c_8)\\
    \c_5 &= -\tfrac12 c_9\\
    \c_6 &= -\tfrac12 c_{10}\\
    \c_7 &= \tfrac14 (c_{11} + i c_{12})\\
    \c_8 &= \tfrac14 (c_{13} + i c_{14})\\
    \c_9 &= -\tfrac12 c_{15}\\
    \c_{10} &= -\tfrac12 c_{16}.\\
  \end{aligned}
\end{equation}

\providecommand{\href}[2]{#2}\begingroup\raggedright\endgroup


\end{document}